 \documentclass[prb,showpacs,showkeys,twocolumn,unsortedaddress]{revtex4}
\usepackage{amsmath,amssymb,epic,eepic}
\usepackage{siunitx}
\usepackage[utf8]{inputenc}
\usepackage{color}
\usepackage{epsfig}
\usepackage{bm}
\usepackage{tikz}
\usepackage{datetime}
\usepackage{natbib}
\usepackage[toc,page]{appendix}

\IfFileExists{srcltx.sty}{\usepackage[active]{srcltx}}

\newcommand{\mi}{\mathrm{i}}

\begin{document}
\title{Unified time-dependent perturbative relations applied to spectroscopy through photo-drag current}
\author{In\`es Safi}
\email{ines.safi@u-psud.fr}

\affiliation{Laboratoire de Physique des Solides (UMR 5802), CNRS-University Paris-Sud/Paris-Saclay,
B\^at. 510, 91405 Orsay, France}

\begin{abstract}
 We develop and exploit an out-of-equilibrium theory, valid in arbitrary dimensions, which does not require initial thermalization. It is perturbative with respect to a weak time-dependent (TD) Hamiltonian term, but is non-perturbative with respect to strong coupling to an electromagnetic environment, or to Coulomb or superconducting correlations. We derive unifying relations between the current generated by coherent radiation or statistical mixture of radiations, superimposed on a dc voltage $V_{dc}$, and the out-of-equilibrium dc current which encodes the effects of interactions. Thus we extend fully the lateral band-transmission picture, thus quantum superposition, to coherent many-body correlated states. This provides methods for a determination of the carrier's charge $q$ free from unknown parameters through the robustness of the Josephson-like frequency. Similar relations we have derived for noise have allowed, recently, to determine the fractional charge in the Fractional Quantum Hall Effect (FQHE) within the Jain series.\citep{christian_photo_2018} The present theory allows for breakdown of inversion symmetry and for asymmetric rates for emission and absorption of radiations. This generates a photo-ratchet effect we exploit to propose a novel method to measure the charge $q$, as well as spectroscopical analysis of the out-of-equilibrium dc current and the third cumulant of non-gaussian statistical radiations. We apply the theory to the Tomonaga-Luttinger Liquid (TLL), showing a counterintuitive feature: a lorentzian pulse superimposed on $V_{dc}$ can reduce the current compared to its dc value, at the same $V_{dc}$, questioning the terminology "photo-assisted". Beyond a charge current, the theory
applies to operators such as spin current in the spin Hall effect, or voltage drop across a phase-slip Josephson junction. 
 \end{abstract}


 \maketitle

Out-of-equilibrium time-dependent (TD) transport offers valuable methods to explore out-of-equilibrium statistical physics, for which novel situations can be monitored, and dynamical properties or characteristic time scales, unveiled by the average current under a DC bias. It has benefited from experimental advances into the high-frequency domain\citep{deblock_06,FF_noise_DCB_altimiras_PRL_2014} or subnanosecond time resolution.\citep{feve_08}  One can distinguish between, on one side, spontaneous generation under a DC bias, such as finite frequency noise, \citep{blanter_buttiker,nazarov_book,schoelkopf_97,deblock_06,deblock_07,sukho_ff_noise,ines_bena,cottet_08,bednorz_13_FF_noise,trauzettel_04,hofheinz_photons_11,zamoum_12,FF_noise_DCB_altimiras_PRL_2014,ines_degiovanni_2016} and, on the other side, phenomena arising from external TD fields, such as pumping, \citep{pump_gossard_Science_99,pump_devoret_PRL_90} mixing\citep{tucker_rev,tomasz_09,thorwart_mix,buttiker_revue_time}
or rectification.\citep{tien_gordon,tucker_rev,buttiker_traversal_time,lesovik_photo,photo_review,ines_eugene} One can as well combine both, such as current noise generated by TD fields.\citep{photo_noise_schoelkopf_98,gabelli_08,glattli_photo_equilibrium,ines_cond_mat,ines_philippe_group,ines_portier_2015} The so-called photo-assisted transport has been interpreted through a key picture, the side-band transmission, expressing quantum superposition of one-particle states.\citep{tien_gordon,tucker_rev,buttiker_traversal_time,lesovik_photo,photo_review} A coherent radiation at a frequency $\Omega_0$ induces inelasticity, as electrons can exchange any number $l$ of photons at the frequency $\Omega_0$. The photo-assisted current is a superposition of replicas of the dc current at an effective dc drive $\omega_J+l\Omega_0$, where:\citep{note_revue}
\begin{equation}\label{josephson_frequency}
\omega_J=\frac{qV_{dc}}{\hbar}
\end{equation}defines a Josephson type frequency, with $q=e$. This picture has some analogy with Shapiro steps in a Josephson Junctions (JJ), where $q=2e$, inspiring the notation. But it has been restricted to: -independent quasiparticles with charge $q=e$, in isolated normal or superconducting junctions (for the quasiparticle current) \citep{tien_gordon,tucker_rev} -a cosine TD voltage, leading to symmetric probabilities of emitting or absorbing $l$ photons -inversion symmetry, with an odd dc current. 
Indeed, it is frequently believed, as expressed by Platero and Aguado in a review paper,\citep{photo_review} that, "for many-body correlated systems, there is no simple picture in terms of side-band transmission". \\
 Here we show that it is still possible to extend such a picture within a unifying theory, restricted by its perturbative nature, but releasing the above restrictions, as it includes simultaneously strongly correlated systems, coherent or statistical mixtures of coherent radiations, and broken symmetries which generate a finite photo-drag current.\citep{ines_cond_mat,ines_eugene}  
It does not seek necessarily an explicit solution, which can be quite involved, but rather relations between the average current under radiations and its average under a dc voltage. The relations unify systems where Coulomb interactions play a fundamental role, such as the Fractional Quantum Hall Effect (FQHE), and those where interactions with an electromagnetic environment lead to phenomena such as the dynamical Coulomb blockade. They can be viewed as perturbative quantum laws for TD transport\citep{odinstov_DCB_88,zaikin_88,ingold_nazarov,girvin_DCB_PRL_90,devoret_prl,zaikin_noise,zaikin,joyez_multi,levy_yeyati,ines_saleur,ines_pierre,baranger_environment} which offer alternatives to classical laws of transport, breaking down. This is in tune with the initial spirit of many-body physics, as exemplified by the phenomenological relations derived by Landau independently on details of Coulomb interactions.\\
The current can refer to a tunneling current between strongly correlated electrodes with mutual Coulomb interactions, a Josephson current in a JJ with a weak Josephson energy and strongly coupled to an electromagnetic environment, or a weak backscattering current between edge states in the FQHE with possible mutual inhomogeneous interactions as well. It can also refer to a spin current between spin Hall edges or a voltage drop across a phase-slip JJ.\citep{photo_josephson_hekking}
 
The perturbative relation for the rectified current extends the side-band transmission picture in terms of the coherent many-body eigenstates of the unperturbed Hamiltonian, exchanging continuous amounts of energy $\hbar\omega'$ with the radiations. This leads to a robust frequency locking through the combination $\omega_J+\omega'$, where $\omega_J$ is given by Eq.(\ref{josephson_frequency}), where $q$ is not necessarily given by $e$; $q$ is fractional in the FQHE or $q=2e$ in a JJ. \\
A first application arises from this robustness: a determination of $q$, linking $V_{dc}$ to $\omega_J$, independently of the detailed microscopic description of the system. This is especially useful for the FQHE at various series of filling factors $\nu$ whose states are not so well understood, and where non-universal features are not easily modelized.\citep{kane_fisher_jain} The best theoretical description has been achieved for $\nu=1/(2n+1)$ with integer $n$, thus within the Laughlin series, where one predicts $q=e/(2n+1)$. If, in addition, lateral confinement is abrupt, the Tomonaga-Luttinger Liquid (TLL) is expected to be appropriate for the edges, the unique model for which photo-assisted transport has been addressed so far.\citep{wen_photo,sassetti_99_photo,photo_lin_fisher,photo_crepieux,photo_TLL_ring_perfetto_13}  \\
An important issue we can also address is the generation of on-demand electrons. Even though many works \citep{feve_07_on_demand,Bocquillon_13_splitter_electrons_demand,parmentier_12_FF_noise_on_demand,note_non_periodic,keeling_06_ivanov,glattli_levitons_nature_13,dubois_minimization_integer} went beyond a cosine TD voltage, they have considered periodic voltages within an independent electron picture. In the limit of a weak current, our theory unifies them with the Tien-Gordon theory, and allows us to address and revisit the minimal excitations in strongly correlated systems or circuits on the one hand, and for non-periodic lorentzian pulses on the other hand.\citep{note_non_periodic}

We will apply our theory to a simple but rich example of ac voltages: a lorentzian pulse superimposed on $V_{dc}$. In particular, in a TLL with repulsive interactions, or in a coherent conductor connected to an ohmic environment,\citep{ines_saleur} the pulse can reduce the current compared to its dc average at $V_{dc}$, even for moderate interactions or resistance. This questions the terminology "photo-assisted", but also the the claim by L. Levitov {\it et al}\citep{keeling_06_ivanov} that the transferred charge is not modified by the pulse. \\
Another application consists into exploiting the photo-drag current to propose spectroscopic methods for the out-of-equilibrium dc current and the third cumulant of a non-gaussian statistical mixture of radiations.

The theory leads to numerous relations\citep{ines_cond_mat,ines_eugene,ines_degiovanni_2016} which have been tested and exploited experimentally,\citep{ines_portier_2015,hofheinz_photons_11,reulet,hofheinz_photons_11,saliha} in particular to determine the fractional charge in the FQHE within the Jain series.\citep{christian_photo_2018} They have been also confirmed by specific theoretical approaches for a periodic TD drive.\citep{photo_josephson_hekking,martin_sassetti_prl_2017}

This is the plan of the paper. Section \ref{sec_model} is devoted to the Hamiltonian and to define the operator $\hat{C}(t)$. Section \ref{sec_average} gives its formal perturbative average under a constant drive, then under TD drives, first at arbitrary frequencies, then at zero-frequency, a limit on which we will focus, and for which special caution is needed. Then we propose some applications, in \ref{sec_applications}, by selecting three profiles of non-periodic TD drives, keeping the generality of the Hamiltonian: -a gaussian pulse -a lorentzian pulse -a non-gaussian statistical mixture of radiations (see appendix \ref{app_periodic} for some periodic profiles). We specify further to a power law dc characteristics in \ref{sec_power_law}, and discuss methods of charge determination in \ref{sec_charge}. We provide the conditions ensuring the validity of the theory in \ref{sec_conditions}, and some examples of unified formal Hamiltonians obeying such conditions in \ref{sec_examples}. 
\section{Model}
\label{sec_model}
\begin{figure}[tb]\begin{center}
\includegraphics[width=6cm]{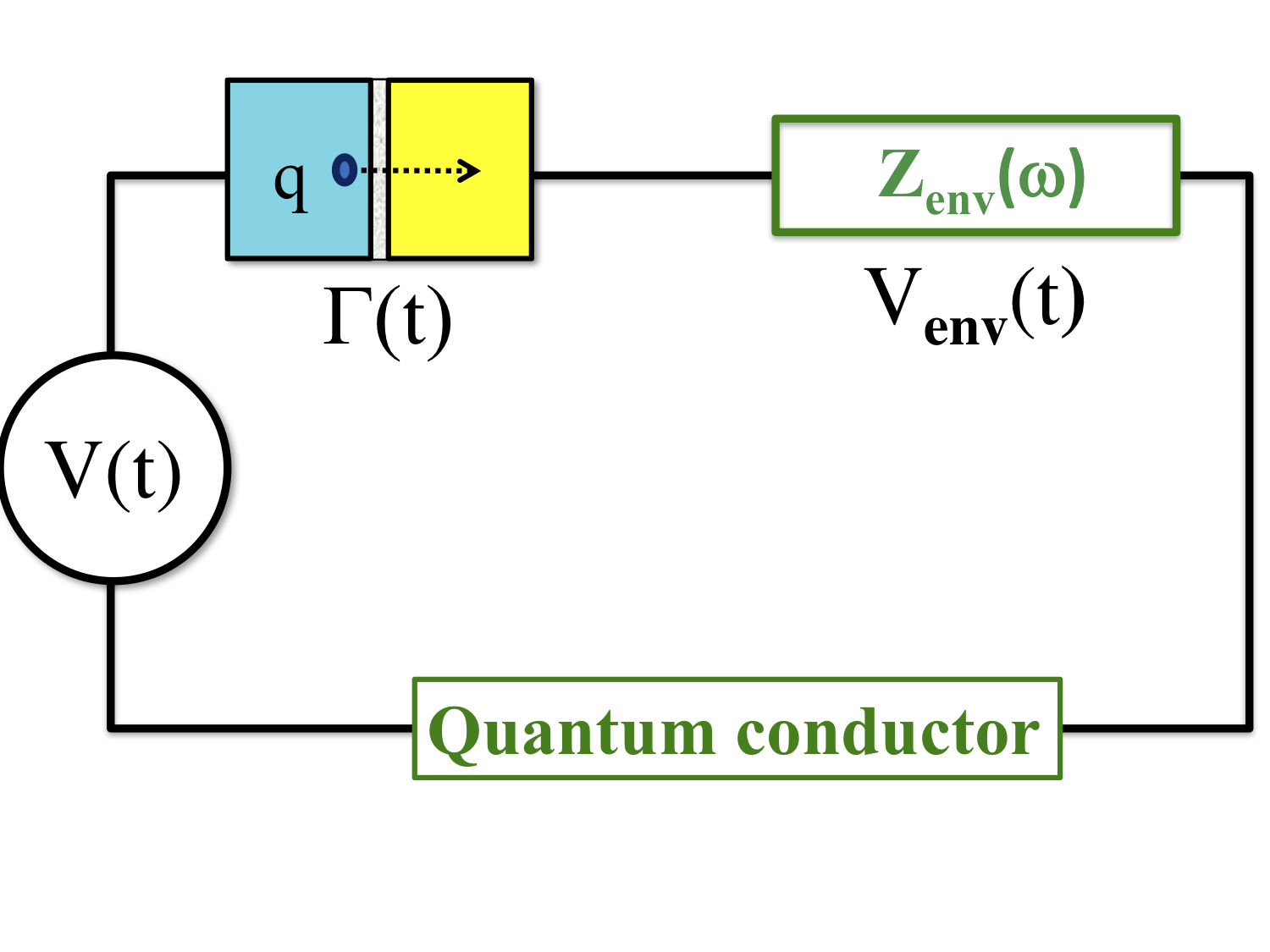}
\caption{\small Family of quantum circuits to which the perturbative theory might apply. A tunneling junction between similar or different electrodes with strong internal and mutual Coulomb interactions, coupled strongly to an electromagnetic environment which could possibly include another quantum conductor. The tunneling amplitude $\Gamma(t)$ depends on time, as well as, possibly, a local magnetic field, gate voltages or voltages across the conductor or the environment, all being encoded into an effective complex function $f(t)$ in Eq.(\ref{Hamiltonian}). The box can refer to a Josephson junction (JJ) with a weak Josephson energy or to a phase-slip JJ.}\label{setup}
\end{center}
\end{figure}

The system we aim to study can contain many entities with mutual couplings, such as electrodes, elements of a quantum circuit or edges/channels. The basic ingredients of the theory are : -a time-independent Hamiltonian $\mathcal{H}_0$ -a perturbing small TD Hamiltonian $\mathcal{H}_{\hat{A}}(t)$, whose TD is factorized through a periodic or non-periodic complex function $f(t)$, independent on position or entities of the system. 
We express it through a weak operator $\hat{A}$ and a free dc drive $\omega_J$: \begin{eqnarray}
	\label{Hamiltonian} 
	\mathcal{H}_{\hat{A}}(t)\!\! &=&\!\!  e^{-i\omega_Jt}{f}(t)\; \hat{A}+ e^{i\omega_Jt}f^*(t)\; \hat{A}^{\dagger}.\\
	\mathcal{H}(t)\!\! &= &\!\! \mathcal{H}_0 + \mathcal{H}_{\hat{A}}(t).
\end{eqnarray} 
 $\mathcal{H}_0$ and $\hat{A}$ are not specified, but have to obey few conditions exposed in section \ref{sec_conditions}. Even though the theory is naturally adapted to a tunneling junction, for which $\hat{A}$ would be the tunneling operator, one can deal here with many features not included within theories of photo-assisted tunneling,\citep{tucker_rev,buttiker_traversal_time,lesovik_photo} for instance (see also Fig.\ref{setup}):
\begin{itemize}
\item Strong Coulomb or superconducting correlations.
\item Strong coupling to an electromagnetic environment.
\item The global Hamiltonian $\mathcal{H}_0$ has not to be split into terms for left and right electrodes, or upper and lower edge states in the Hall regime, thus can include mutual Coulomb interactions, possibly inhomogeneous.
\item  $\mathcal{H}_{\hat{A}}(t)$ could describe non-local processes with respect to entities or space, such as fixed/random tunneling/backscattering processes due to extended impurities or Coulomb interactions (see Eq.(\ref{A(x)_0})). \end{itemize}    
 We can assume ${f}(t=0)\!=\!1$ by renormalizing $\hat{A}$, thus the stationary regime corresponds to $f(t)=1$. Let's now introduce:\begin{eqnarray}
{f}(t)&=&|f(t)|e^{-i\varphi(t)}\\
W_{ac}(t)&=&\partial_t\varphi(t)\label{f_phase}\\
W(t)&=&\omega_J+ W_{ac}(t),\label{total_W}
\end{eqnarray} where any constant in $W_{ac}(t)$ is implicitly translated to $\omega_J$, so that $W_{ac}(t)$ becomes integrable. Our aim is to express, to second order with respect to $\hat{A}$, the average of the generalized force derived from $\mathcal{H}_{\hat{A}}(t)$:
\begin{equation}
	\label{eq:current}
	i\hbar\hat{C}(t)\! =\frac{\delta\mathcal{H}_{\hat{A}}(t)}{\delta\varphi(t)}\!=
		e^{-i\omega_Jt}{f}(t)\;\hat{A}-e^{i\omega_Jt}f^*(t)\; 
\hat{A}^{\dagger}\,.
\end{equation}  
The observable associated with $\hat{C}(t)$ depends on the model. It could be charge or spin currents in Quantum Hall, superconducting or magnetic conductors, or a voltage operator in phase-slip JJs. It could also provide a weak correction, induced for instance by weak impurities described by $\mathcal{H}_{\hat{A}}$, to a finite average current or voltage in presence of $\mathcal{H}_{0}$ only. 
We define average with respect to an initial density matrix $\hat{\rho}_0$ at time=$-\infty$: $<...>=Tr (\hat{\rho}_0 ...)/Tr{\hat{\rho}_0}$.\\
 \begin{figure}[tb]\begin{center}
\includegraphics[width=6cm]{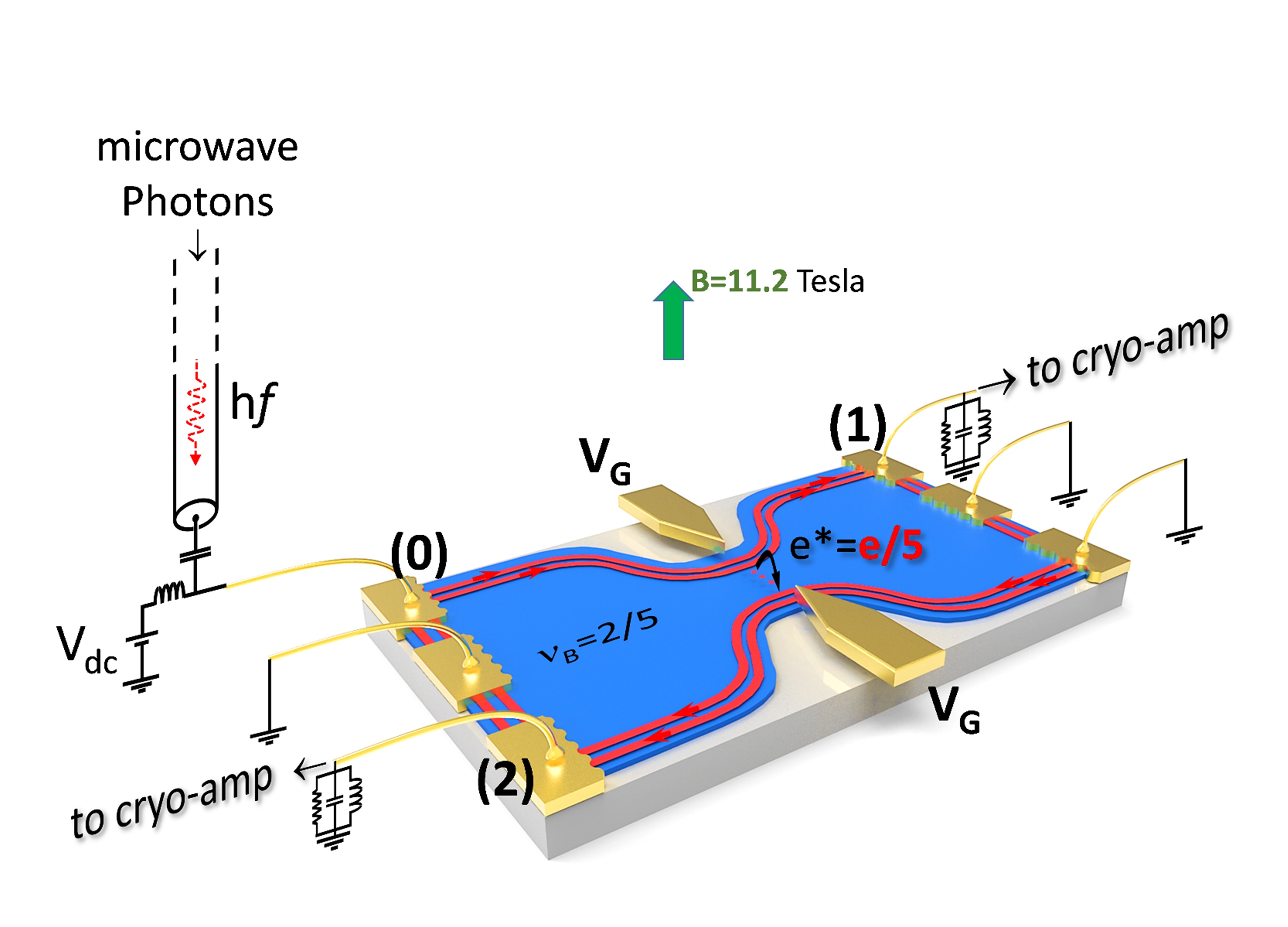}
\caption{\small  A quantum point contact (QPC) in the FQHE at $\nu= 2/5$. This corresponds to the first Jain filling factor series, $\nu= p/(2p +1)$ with integer number p, where one expects $q = e/(2p+1)$.
To experimentally determine $q$,\citep{christian_photo_2018} a RF excitation at frequency f is added to the dc voltage $V_{dc}$ applied to contact (0). This contact injects carriers at the bulk filling factor $\nu_B = 2/5$ into two chiral fractional edge modes (red lines). The carriers are partitioned by the QPC into transmitted and reflected current measured at contacts (1) and (2), from which the current fluctuations  are recorded and their cross-correlation performed. The charge $q = e/3$ has been determined through the Josephson  relation $f_J = qV_{dc}/\hbar$. The charge q = e/5 has been also measured for a weak backscattering of the inner edge. The charge determination is based on the robust relation for the photo-assisted noise within our perturbative theory,\citep{ines_cond_mat} the first to address photo-assisted transport beyond the TLL description, at simple fractions. As it is independent on the precise model for the edges, all series of filling factors, mutual Coulomb interactions, inhomogeneities at the QPC or edge reconstruction could in principle be also addressed.}\label{setup_hall}
\end{center}
\end{figure}
The present model forms a particular family within the domain of validity of the non-perturbative theory in Ref.\citep{ines_philippe_group}, which is almost free of any conditions on a TD Hamiltonian. Being also non-perturbative with respect to strong correlations or coupling to an electromagnetic environment, the main restrictions on the present theory resides into its perturbative nature and its restriction to a unique operator $\hat{C}(t)$, in Eq.(\ref{eq:current}), deriving from $\mathcal{H}_{\hat{A}}(t)$. But paying such a "price" has the advantage to offer analysis of its weak average. Still both theories\citep{ines_philippe_group,ines_cond_mat,ines_eugene} have in common that they provide quantum laws for TD out-of-equilibrium transport which are respectively exact and perturbative. \section{Out-of-equilibrium averages of $\hat{C}(t)$}
\label{sec_average}
\subsection{A constant drive}
We address first the stationary regime, under a constant drive $\omega_J$, thus we let ${f}(t)=1$ in Eq.(\ref{Hamiltonian}).
We denote  the average of $\hat{C}(t)$, or, more precisely, its second-order term with respect to $\hat{A}$, by:
 \begin{equation}\label{Fdc} 
C_{dc}(\omega_J)=<\!\!\hat{C}_\mathcal{H}(t)\!\!>_{{f}(t)=1},
\end{equation} where the subscript ${\mathcal{H}}$ refers to Heisenberg representation with respect to the total Hamiltonian ${\mathcal{H}}(t)$ in Eq.(\ref{Hamiltonian}). Notice that $\omega_J$ is not fixed to its value in Eq.(\ref{Hamiltonian}) when combined with $f(t)$, but plays rather the role of a continuous variable.  \\
Though we don't restrict $\hat{\rho}_0$ to a standard thermal distribution, we assume nonetheless that :
\begin{equation}\label{condition_matrix}
[\hat{\rho}_0,\mathcal{H}_0]=0,
\end{equation}
i.e. that $\hat{\rho}_0$ is diagonal in the basis of eigenstates of $\mathcal{H}_0$, denoted $|n\!>$ with energy $E_n$. Thus $\hat{\rho}_0$ is stationary in absence of $\mathcal{H}_{\hat{A}}$ (Eq.(\ref{Hamiltonian})), which ensures, partly, that equation (\ref{Fdc}) is stationary. 
A spectral decomposition of $C_{dc}(\omega_J)$ in this basis, letting $\rho_n=<n|\hat{\rho}_0|n>$ and $A_{m,n}=<m|\hat{A}|n>$, reads:
\begin{equation}\label{spectral_F}
C_{dc}(\omega_J)=\sum_{n,m}(\rho_n-\rho_m)|\hat{A}_{m,n}|^2 \delta(E_n-E_m-\hbar \omega_J).
\end{equation} 
 We don't require inversion symmetry, thus one might have $C_{dc}(\omega_J)+C_{dc}(-\omega_J)\neq 0$.\\
 Let us discuss now the limit of vanishing $\omega_J$, if located within the perturbative regime. $C_{dc}(\omega_J)$ might contain a singularity of the form $\delta(\omega_J)$ due to degenerate states. Even though this would be interesting to study, we simplify the discussion by assuming further that $\rho_n$ is entirely determined by the energy $E_n$ of $|n\!>$: 
\begin{equation}\label{supplementary_condition}
E_n=E_m\Rightarrow \rho_n=\rho_m. 
\end{equation}Therefore, using Eq.(\ref{spectral_F}), we get :
 \begin{equation} \label{vanishing}
 C_{dc}(\omega_J=0)=0.
\end{equation} This can be interpreted, when $\hat{C}$ refers to a current operator, as a negligible super-current. This is ensured for instance in a JJ coupled to a dissipative environment, where, at energies below the gap, $\hat{C}(t)$ refers to a weak Josephson current, or in SIN junctions at energies above the gap, where $\hat{C}(t)$ refers to the quasiparticle current, $q=e$, and the super-current is made negligible by applying a magnetic field.

Let us now express the average of $\hat{C}(t)$ under TD drive:
\begin{equation}\label{Fac_t}
C_{f}(\omega_J;t)=<\!\!\hat{C}_\mathcal{H}(t)\!\!>_{f(t)\neq 1},
\end{equation}
whose functional dependence on the complex function $f(t)$ is recalled through the subscript $f$. 
The conditions given in section \ref{sec_conditions} allow us to express $C_{f}(\omega_J;t)$ to second order with respect to $\hat{A}$, or its Fourier transform at a finite frequency $\omega$, $C_{f}(\omega_J;\omega)$. Indeed, even though  $C_{f}(\omega_J;\omega)$ and $C_{dc}(\omega_J)$ in Eq.(\ref{Fdc}) are out-of-equilibrium observables, they can be expressed, owing to the perturbative approach, in terms of an equilibrium retarded correlator (given by. Eq.(\ref{I_DC})) forming the bridge between them. 
Then, without knowledge of the many-body eigenstates nor of the initial diagonal elements of $\hat{\rho}_0$, we show that the dc average $C_{dc}(\omega_J)$ is sufficient to incorporate features of the Hamiltonian, and to determine $C_{f}(\omega_J;\omega)$:
\begin{eqnarray}\label{Iomega_bis}
C_{f}(\omega_J;\omega)\!\! &=&\!\! \mi\int_{-\infty}^{+\infty}\!\!\! \int_{-\infty}^{+\infty}\!\!\!{d\omega'} d\omega" f^*(\omega'-\omega/2)
 f(\omega'+\omega/2)\nonumber\\&&\frac{(\omega/2+\mi\delta)C_{dc}(\omega_J+\omega"+\omega')}{\omega"^2-(\omega/2+\mi\delta)^2},
\end{eqnarray} 
with $ f(\omega')$ the Fourier transform of the complex $f(t)$. 
This is a first central relation of the paper, which we keep at this formal level. \\
We rather focus on the limit of low frequencies $\omega$, the most accessible experimentally, which obeys the second central relation of the paper (see also appendix \ref{app_derivation}):  
\begin{equation}\label{zero_frequency_w}
C_{f}(\omega_J;\omega\rightarrow 0)\!\! = \int_{-\infty}^{+\infty}\!\!\frac{d\omega'}{2\pi} f^*(\omega'+\omega) f(\omega') C_{dc}(\omega_J+\omega').
\end{equation}
We have kept the small frequency $\omega$ on the r.h.s. because $f(\omega')$, thus $C_{f}(\omega_J;\omega)$, are not necessarily regular at zero frequency.\\
 We have to stress that we don't compute the function $C_{dc}(\omega_J)$, which would provide informations on $\mathcal{H}(t)$, as is the usual task of theoretical studies. We are indeed showing that we don't gain more informations on the underlying Hamiltonian by studying $C_{f}$ compared to $C_{dc}$, which is, in some sense, disappointing. Indeed, the relation in Eq.(\ref{zero_frequency}) holds even in case $C_{dc}(\omega_J)$ becomes inaccessible, due to the complexity of $\mathcal{H}(t)$. It will also lead us to very interesting observations and applications throughout the paper, which arise, roughly, along three ways, depending on whether the unknown function to be determined using the two others is: either $C_{dc}(\omega_J)$, $f(\omega)$ or $C_{f}(\omega_J;\omega)$. \\
The integral on the r.h.s. of Eq.(\ref{zero_frequency_w}) is assumed implicitly to run over frequencies within the domain of validity of the perturbative theory. Its convergence criteria depends on each specific model and profile of $f(t)$, but can be facilitated by a finite measurement time $T_0$, as we discuss now. 
\subsubsection*{Zero-frequency measurement}
Here, we address a feature we have not yet clarified in our previous related works, which is relevant to a non-periodic $f(t)$ for which $f(\omega)$ contains a delta function: 
 \begin{equation}\label{dec_f}
 f(\omega)=2\pi f_{dc}\delta(\omega)+f_{ac}(\omega).
 \end{equation} 
$f_{dc}$ is a complex number and $f_{ac}(\omega)$ is regular, possibly finite, at $\omega=0$. This is the case when $W(t)$ (see Eq.(\ref{f_phase})) is formed by a single lorentzian pulse (see section \ref{sec_applications}), or if $f(t)=f_{dc}+f_{ac}(t)$, with $f_{ac}(t)$ integrable. Given a function $g(t)=g_{dc}+g_{ac}(t)$, with $g_{ac}(t)$ integrable, one expects the dc value to induce averaging over a measurement time $T_0$:
\begin{equation}\label{g_meas}
g^{(0)}=\frac{1}{T_0}\int_{-\frac{T_0}2}^{\frac{T_0}2}g(t)=g_{dc}+\frac{1}{T_0}g_{ac}(\omega=0).
\end{equation}
If $g(t)$ is periodic, $T_0$ would be simply its period, and one is free to choose $g_{dc}=0$.
We have implicitly made a similar decomposition of $W(t)$ in Eq.(\ref{total_W}), so that its effective dc component is given, using Eq.(\ref{g_meas}):  
\begin{equation}\label{W0}
W^{(0)}=\omega_{J}+\frac{1}{T_0}W_{ac}(\omega=0).
\end{equation}
Following again Eq.(\ref{g_meas}) to define the zero-frequency average $C_{f}^{(0)}(\omega_J)$ in terms of $C_f(\omega_J;t)$, in Eq.(\ref{Fac_t}), we can infer it from Eq.(\ref{zero_frequency_w}):\citep{note_dimension}
\begin{eqnarray}\label{zero_frequency}
C^{(0)}_{f}\!(\omega_J)&=&\left(|f_{dc}|^2+\frac 2{T_0}Re \left[f_{dc} f_{ac}^*(0)\right]\right)C_{dc}(\omega_J)\nonumber\\&&+\;C_{f_{ac}}^{(0)}\!(\omega_J),
\end{eqnarray}
 where:
 \begin{equation}\label{zero_frequency_ac}
C_{f_{ac}}^{(0)}(\omega_J)\!\! =\!\! \int_{-\infty}^{+\infty}\!\!\frac{d\omega'}{2\pi} p(\omega') C_{dc}(\omega_J+\omega'),
\end{equation} and:
 \begin{equation}\label{p}
p(\omega')=\frac{1}{T_0}\int_{-\frac{T_0}2}^{\frac{T_0}2}dt\int_{-\infty}^{\infty} dt' e^{i\omega'(t-t')}f_{ac}(t)f_{ac}^*(t').
\end{equation}
 Notice that for $f_{dc}=0$, equation (\ref{zero_frequency}) reduces to $C^{(0)}_{f}\!(\omega_J)=C_{f_{ac}}^{(0)}\!(\omega_J)$. $p(\omega')$ in Eq.(\ref{p}) has a precise probabilistic meaning only for $|f_{ac}(t)|=1$, for which $\int d\omega' p(\omega')=1$. But even if $|f_{ac}(t)|$ depends on time, $p(\omega')$ can still be viewed as a transfer rate for emitting (resp. absorbing) an energy $\hbar\omega'$ for positive (resp. negative) $\omega'$, generally different as $p(\omega')$ is not necessarily even.\\ Indeed, even when the one-electron picture is inappropriate and $\hat{C}(t)$ does not refer to a current, we can interpret $C_{f_{ac}}^{(0)}\!(\omega_J)$ in Eq.(\ref{zero_frequency_ac}) (the contribution due to $f_{ac}$ in Eq.(\ref{dec_f})) once we inject the spectral decomposition of $C_{dc}(\omega_J)$ in Eq.(\ref{spectral_F}). We can extend simultaneously two pictures: -lateral side-band transmission in terms of global many-body eigenstates $|n>$ of $\mathcal{H}_0$ -dynamical Coulomb blockade, the non-periodic radiations acting as a classical electromagnetic environment. In addition to the energy furnished by the dc drive $\hbar\omega_J$, there are transitions between many-body eigenstates $|n>$ by exchanging energy $\hbar\omega'$ with $f_{ac}(t)$, leading to an effective dc drive $\omega_J+\omega'$. $C_{f_{ac}}^{(0)}(\omega_J)$ in Eq.(\ref{zero_frequency_ac}) is obtained by integrating over $\omega'$ the dc average $C_{dc}(\omega_J+\omega')$, modulated by the transfer rate $p(\omega')$. This is a quantum superposition of many-body eigenstates, global quantum coherence being maintained with Coulomb interactions or dissipation due to an ohmic environment. 
The frequency locking through the combination $\omega_J+\omega'$ on the r.h.s. of Eq.(\ref{zero_frequency_ac}), thus the way the Josephson-type frequency $\omega_J$ intervenes, is independent on details of the Hamiltonian (\ref{Hamiltonian}) and the nature of the operator $\hat{C}(t)$. 
 \subsection{Photo-ratchet effect}
 Let us now discuss the limit when $\omega_J=0$, if located within the perturbative domain.\citep{note_total_0}The expression for the zero-frequency average in Eq.(\ref{zero_frequency}) is now independent on whether $f(\omega)$ is singular or not, and using Eq.(\ref{vanishing}), it reduces to Eq.(\ref{zero_frequency_ac}):
\begin{equation}C_{f}^{(0)}(\omega_J=0)=C_{f_{ac}}^{(0)}(\omega_J=0).
\end{equation}
Now we observe that $C_{f}^{(0)}(\omega_J=0)=0$ whenever one has simultaneously:
\begin{eqnarray} 
C_{dc}(\omega_J)&=&-C_{dc}(-\omega_J).\label{inversion}\\
p(\omega)&=&p(-\omega).\label{symmetric_p}
\end{eqnarray} As soon as one of those two symmetries is broken, a photo-drag average is obtained, 
 $$C_{f}^{(0)}(\omega_J=0)\neq 0.$$ This photo-ratchet effect will be exploited later on to propose some probing methods. Here we will choose asymmetric $p(\omega)$, while the dc average respects the inversion symmetry in Eq.(\ref{inversion}), whose breakdown is reported to a separate work.
  \subsection{Case of a current operator}
To obtain a charge current operator, one needs a renormalizing charge $q'$, possibly effective and depending on $\mathcal{H}_0$:
  \begin{equation}\label{current}\hat{I}(t)=q'\hat{C}(t).
\end{equation} 
One needs also a renormalizing charge $q$ of the voltage, possibly different from $q'$ (see Eq.(\ref{f_phase},\ref{total_W})): 
\begin{eqnarray}\label{equivalence}\hbar\omega_J&=&qV_{dc}.\\
\hbar W_{ac}(t)&=&q V_{ac}(t).
\end{eqnarray} 
We also introduce the total voltage: 
\begin{equation}\label{V_dc_ac}
V(t)=V_{dc}+V_{ac}(t).
\end{equation}
As in Eq.(\ref{W0}), $V^{(0)}$ refers to the average of $V(t)$ over one period, and to Eq.(\ref{g_meas}) for a non-periodic $V(t)$:
\begin{equation}\label{V0}
 V^{(0)}=V_{dc}+\frac{1}{T_0}V(\omega=0).
 \end{equation}
The relation in Eq.(\ref{zero_frequency}), replacing $C$ by $I$ on both sides, provides $I_{f}^{(0)}(\omega_J)$ in terms of $I_{dc}(\omega_J)$, which is determined in a non-trivial way by both $|f(t)|$ and $V_{ac}(t)$. \\
$I_{f}^{(0)}(\omega_J)$ could also give (up to a possible renormalisation) a perturbative correction induced by $\mathcal{H}_{\hat{A}}(t)$ to a non-vanishing average current in presence of $\mathcal{H}_0$. This is the case when $\hat{I}(t)$ corresponds to the current operator in one branch of a quantum circuit,\citep{photo_grabert_current} or to the weak backscattering current induced by impurities (as in section \ref{sec_power_law}).\\   Using Eq.(\ref{zero_frequency}),  one can also obtain the expression of the differential photo-conductance $G_f(\omega_J)=dI_f^{(0)}(\omega_J)/dV_{dc}$ in terms of the differential dc conductance: $G_{dc}(\omega_J)={dI_{dc}(\omega_J)}/{dV_{dc}}$ (see appendix \ref{app_derivation}).
Notice that for a non-linear dc current, $G_f(\omega_J)$ is different from differentials of $I_f^{(0)}(\omega_J)$ with respect to $V_{ac}(\omega\neq 0)$.\citep{ines_cond_mat,ines_eugene} Indeed, the validity of the perturbation has to be expressed by the weakness of $G_{dc}(\omega_J)$. If it provides the correction to a finite conductance $G_0(\omega_J)$ in presence of $\mathcal{H}_0$ only, one needs: $|G_{dc}(\omega_J)|\ll |G_0(\omega_J)|$.\\
We address the case of a linear dc current, thus $G_{dc}(\omega_J)=G_{dc}$ constant, in appendix \ref{app_linear}. We show that $I_{f}^{(0)}(\omega_J)$ is determined in a non-trivial way by $|f(t)|$ and $V(t)$, but vanishes whenever $V(t)=0$ (see ). It is only when $|f(t)|=1$ that $I_{f}^{(0)}(\omega_J)$ is determined only by $V^{(0)}$ in Eq.(\ref{V0}), and that $G_f=G_{dc}$.

Let's now comment the case of a periodic $f(t)$ with period $T_0$ (see Ref.[\citep{ines_eugene}] and appendix \ref{app_periodic}). Then $p(\omega)$ takes discrete values $p_l$, the rates at which many-body eigenstates of $\mathcal{H}_0$ exchange $l$ photons of frequency $\Omega_0=2\pi/T_0$ :
\begin{equation}\label{p_periodic}
p(\omega)=\sum_{l=-\infty}^{\infty} p_l\delta(\omega-l\Omega_0).
\end{equation}
The integral in Eq.(\ref{zero_frequency}) reduces now to a discrete sum:\citep{ines_eugene}
\begin{eqnarray}\label{zero-frequency-periodic_charge}
I_{f}^{(0)}(\omega_{J})
&=&\!\! \sum_{l=-\infty}^{\infty}\! p_l \;I_{dc}(\omega_J+l\Omega_0).
\end{eqnarray}
We can express similarly the differential photo-conductance in Eq.(\ref{photo_conductance}) in terms of $G_{dc}$ (see appendix \ref{app_periodic}).\\
 Let's now specify to the frequently adopted profile: $W_{ac}(t)=W_{\Omega_0}\cos \Omega_0 t$ with a constant $|f(t)|=1$, so that $p_l$ are given by Bessel functions $J_l$ of the first kind:
\begin{equation}\label{bessel}
p_l=\left|J_{l}\left(\frac{W_{\Omega_0}}{\Omega_0}\right)\right|^2.
\end{equation} 
The relation in Eq.(\ref{zero-frequency-periodic_charge}), now similar to the Tien-Gordon formula for photo-assisted tunneling current\citep{tien_gordon,tucker_rev} at $\omega_{J}=eV_{dc}/\hbar$, unifies numerous works with explicit models and derivations, either for independent carriers\citep{tien_gordon,tucker_rev,buttiker_traversal_time,lesovik_photo} or within the TLL model.\citep{wen_photo,sassetti_99_photo,photo_lin_fisher,photo_crepieux,photo_TLL_ring_perfetto_13} It is now extended to a much larger domain of validity. The charges $q,q'$ entering in Eqs.(\ref{equivalence},\ref{current}), thus in $\omega_{J}=qV_{dc}/\hbar$, can be different from $e$, for instance fractional in the FQHE and $2e$ in JJs, and the dc current is not necessarily odd, thus we can obtain a photo-ratchet effect, contrary to Tien-Gordon theory.

 \section{Selected profiles of $f(t)$: spectroscopic methods}
\label{sec_applications}
For all applications, we consider, for familiarity and comparison to related works, the case $\hat{C}(t)$ in Eq.(\ref{eq:current}) refers to a current operator $\hat{I}(t)$, thus we exploit the relation in Eq.(\ref{zero_frequency}) replacing $C$ by $I$. Without specifying the Hamiltonian terms in Eq.(\ref{Hamiltonian}), thus keeping undetermined the dc current $I_{dc}(\omega_J)$, we choose three non-periodic profiles of $f(t)$: 
\begin{itemize}
\item $W_{ac}(t)$ is a lorentzian pulse, and $|f(t)|=1$ constant. 
\item $|f(t)|$ is a gaussian pulse, with a constant phase, thus $W_{ac}(t)=0$. 
\item $W_{ac}(t)$ is a non-gaussian statistical mixture of radiations. 
\end{itemize}
We obtain formal results, which will be detailed (for a lorentzian and gaussian pulses) in the next section, when the dc characteristics is a power law. We show also, in appendix \ref{app_periodic}, how periodic pulses generate a Josephson-type oscillation, without need to any superconducting correlations. 
 \subsection{A gaussian pulse}
To illustrate some advantages of non-periodicity, we choose $f(t)$ such that $p(\omega)$ in Eq.(\ref{p}) is peaked around a frequency $\omega_p$. We also let $f_{dc}=0$ in Eq.(\ref{dec_f}), thus $f(\omega)=f_{ac}(\omega)$. For instance, if $f_{ac}(t)$ is real (or at least $\varphi(t)$ constant, i.e. $W_{ac}(t)=0$) and gaussian with a large width in time, using a TD gate voltage and a gaussian filter, the transfer rate in Eq.(\ref{p}) is also gaussian with a small width $\sigma$ (compared to other frequency scales):
\begin{equation}\label{f_peak}
p(\omega)=\frac{1}{\sqrt {2\pi}\sigma}e^{-\frac{(\omega-\omega_p)^2}{2\sigma^2}}.
\end{equation}Then, using Eq.(\ref{zero_frequency_ac}), the induced current is related in a simple way to the dc current:
\begin{equation}\label{zero-frequency-current_peak}
I_{f}^{(0)}(\omega_J)\!=\!I_{dc}(\omega_J+\omega_p).
\end{equation}
Again, even when the dc current $I_{dc}(\omega_J)$ is too complicated to compute, this relation holds without a precise knowledge of Hamiltonian terms in Eq.(\ref{Hamiltonian}). It offers a promising method to infer the dc out-of-equilibrium current from a measurement of $I_{f}^{(0)}(\omega_J)$. This is especially convenient for a thermal initial distribution, where the out-of-equilibrium domain corresponds often to high dc voltages, $\hbar\omega_J\gg k_B T$, which can nonetheless cause heating. Thus, in Eq.(\ref{zero-frequency-current_peak}), one could keep $\hbar\omega_J\ll k_B T$, but choose $\omega_p\gg k_B T$. Taking further a vanishing dc voltage, one gets the photo-drag current,
 \begin{equation}\label{zero-frequency-current_peak_zero}
I_{f}^{(0)}(\omega_J=0)\!\! =\!\! I_{dc}(\omega_p),
\end{equation}
 whose measurement at $\omega_p\gg k_B T/\hbar$ provides $I_{dc}(\omega_J=\omega_p)$. One needs nonetheless to vary $\omega_p$ in order to explore a large interval of dc voltages.
\subsection{A lorentzian pulse}
For $W_{ac}(t)$ in Eqs.(\ref{f_phase},\ref{total_W}) formed by an arbitrary series of periodic or non-periodic lorentzian pulses, we can, in principle, obtain $I_{f}^{0)}(\omega_J)$ from Eq.(\ref{zero_frequency}) (replacing $C$ by $I$). If we specify to $|f(t)|=1$, thus we let:
 \begin{equation}\label{f_phi}
 f(t)=e^{-i\varphi(t)},
 \end{equation} where $\partial_t\varphi(t)={W}_{ac}(t)$, the contribution of either positive or negative $\omega'$ to the integral in Eq.(\ref{zero_frequency}) vanishes. We can also show, for a non-periodic $W_{ac}(t)$, that $W_{ac}(\omega=0)$ is still an integer multiple of $2\pi$, even when $q$ is fractional, as noticed in Ref.\citep{levitons_glattli_review_2017} for a periodic $W_{ac}(t)$.\\ For simplicity, we specify here
 to a single lorentzian pulse centered around a time $t_1$, with a width $\tau_1$:
\begin{equation}\label{lorentzian}
{W}_{ac}(t)=\frac {2\tau_1}{(t-t_1)^2+\tau_1^2}.
\end{equation}
Now the dc component of ${W}_{ac}(t)$ is given by: $\int_{-\infty}^{+\infty} dt \;W_{ac}(t)=W_{ac}(\omega=0)=2\pi$, and, as we add a free dc drive $\omega_J$ to $W_{ac}(t)$ (see Eqs.(\ref{Hamiltonian},\ref{total_W})), the dc component in Eq.(\ref{W0}) becomes:
\begin{equation}
W^{(0)}=\omega_J+\Omega_0,
\end{equation}
where $\Omega_0=2\pi/T_0$. The Fourier transform of Eq.(\ref{f_phi}), given Eq.(\ref{lorentzian}), is:
\begin{equation}\label{f_ac_lorentzian}
{f(\omega)}=2\pi\delta(\omega)+f_{ac}(\omega),
\end{equation}
where:
\begin{equation}\label{f_ac_lorentzian_bis}
{f_{ac}(\omega)}=-4\pi\tau_1\theta(\omega)e^{-\omega(\tau_1-it_1)},
\end{equation}
 $\theta$ being the Heaviside function. 
Using Eq.(\ref{zero_frequency}) with $f_{dc}=1$, we find:
\begin{equation}\label{F_z_lorentzian}
 {I}_{f}^{(0)}(\omega_J)= \left(1-2\tau_1\Omega_0\right)I_{dc}(\omega_J)+{I}_{f_{ac}}(\omega_J),
 \end{equation}
 where:
 \begin{equation}\label{F_z_lorentzian_ac}
 {I}_{f_{ac}}(\omega_J)=4\Omega_0\tau_1^2\int_0^{\infty} {d\omega'} e^{-2\omega'\tau_1} I_{dc}(\omega_J+\omega').
 \end{equation}
 We recall that this relation is independent on the Hamiltonian terms in Eq.(\ref{Hamiltonian}) and the initial diagonal $\hat{\rho}_0$, which intervene only through $I_{dc}(\omega_J)$. 

Let's now take the limit $\omega_J=0$ in Eq.(\ref{F_z_lorentzian}):
  \begin{equation}\label{F_z_lorentzian_current}
{I}_{f}^{(0)}(\omega_J=0)=4\tau_1^2\int_0^{\infty} \frac{d\omega'} {2\pi} e^{-2\omega'\tau_1} I_{dc}(\omega'),
 \end{equation}which can be interpreted as follows. A dc voltage $V_{dc}=\hbar\omega'/q$ injects elementary charges $q$ with period $2\pi/\omega'$, the integral on the r.h.s. is dominated by the contribution of dc voltages for which this period is smaller than the width of the pulse: $2\pi/\omega'<2\tau_1$.
 Notice that it is only only when $I_{dc}(\omega')=G_{dc}\hbar\omega'/q$ is linear (appendix \ref{app_linear}) that we recover the result by L. Levitov {\it et al},\citep{keeling_06_ivanov} (who adopt $V_{dc}=0$), mainly that ${I}_{f}^{(0)}(\omega_J=0)$ in Eq.(\ref{F_z_lorentzian_current}) is determined only by the area of the pulse $V_{ac}(\omega=0)$, given here by $e$.
\subsection{Non-gaussian statistical mixture of radiations: probing the third cumulant}
Here we consider the case where $f(t)$ is generated by a statistical mixture of coherent radiations. This arises, for instance, in the classical regime of an electromagnetic environment, not included in the Hamiltonian to avoid redundancy. For simplicity, we assume that $f_{dc}=0$ in Eq.(\ref{dec_f}), $|f_{ac}(t)|=1$ and $\varphi (t)$ is a fluctuating phase with a random distribution $\mathcal{D}(\varphi)$. Therefore, one needs to perform the average of Eq.(\ref{zero_frequency_ac}) over $\mathcal{D}$, denoted by $<...>_{\mathcal{D}}$, and included in the definition of the rate transfer $p(\omega)$, Eq.(\ref{p}):
\begin{equation}\label{p_average}
p(\omega)=\int_{-T_0/2}^{T_0/2}\frac{dt}{T_0}\int_{-\infty}^{\infty}dt' e^{i\omega(t-t')} <e^{i\varphi(t)-i\varphi(t')}>_{\mathcal{D}}.
\end{equation}
One can write the average on the r.h.s. as exponential of cumulants of $\varphi(t)$. \begin{eqnarray}
J_m(t,t')&=&\frac {i^m} {{m}\mathpunct{!}} <(\varphi(t)-\varphi(t'))^m>_{\mathcal{D}}.\label{Jm}
\end{eqnarray}
The expression, not given here, is simpler compared to a quantum phase operator,\citep{sukho_detection} as $\varphi(t)$ is classical. \\
Consider first the case when ${\mathcal{D}}$ is symmetric : 
\begin{equation}\label{Symmetry_P}
\mathcal{D}(\varphi)=\mathcal{D}(-\varphi),
\end{equation} 
in which case only cumulants with even $m$ survive. Changing $\varphi\rightarrow-\varphi$ in Eq.(\ref{p_average}) amounts to permute $t$ and $t'$ on the r.h.s. of Eq.(\ref{p_average}), thus one has $p(\omega)=p(-\omega)$. This holds in particular whenever $\mathcal{D}$ is a gaussian functional of $\varphi(t)$:
\begin{equation}\label{p_average_gaussian}
p(\omega)=\int_{-T_0/2}^{T_0/2}\frac{dt}{T_0}\int_{-\infty}^{\infty}dt' e^{i\omega(t-t')} e^{-J_2(t,t')},
\end{equation}
Consider secondly the case when the symmetry in Eq.(\ref{Symmetry_P}) is broken, leading to finite $J_m$ with odd $m$, thus an asymmetry of Eq.(\ref{p_average}): $p(\omega)\neq p(-\omega)$.  
 Let us further assume that $\varphi(t)=\lambda \phi(t)$,
 where $\lambda$ is a small (coupling) parameter and $\phi(t)$ a fluctuating field. Expanding Eq.(\ref{p_average}) to order $\lambda^3$ yields:
 \begin{equation}\label{p_average_non_gaussian}
 p(\omega)=-J_2(\omega)+ J_3(\omega),
\end{equation}
where $J_2,J_3$ are the second and third cumulant of $\varphi(t)$, given by Eq.(\ref{Jm}), and:
\begin{equation}
J_m(\omega)=\int_{-T_0/2}^{T_0/2}\frac{dt}{T_0}\int_{-\infty}^{\infty}dt' e^{i\omega(t-t')} J_m(t,t'),
\end{equation}
We see that $J_2(t,t')=J_2(t',t)$, while $J_3(t',t)=-J_3(t,t')$, so that:
\begin{eqnarray}\label{parity_J}
J_2(\omega)&=&J_2(-\omega)\nonumber\\
J_3(\omega)&=&-J_3(-\omega).\label{J_m}
\end{eqnarray}
The relation in Eq.(\ref{zero_frequency_ac}) opens the path to probe the third cumulant $J_3(\omega)$ by measuring $I^{(0)}_f(\omega_J)$.\\
One way to proceed would be to let $\omega_J=0$, thus to exploit the photo-ratchet effect. It is then convenient to choose a dc current which respects inversion symmetry, $I_{dc}(\omega)=-I_{dc}(-\omega)$. Using Eqs.(\ref{zero_frequency_ac},\ref{p_average_non_gaussian},\ref{parity_J}), we obtain a relation between the photo-drag current and the third cumulant:
 \begin{equation}\label{photo_J3}
 I_{f}^{(0)}(\omega_J=0)=2 \int_{0}^{+\infty}\!\!\frac{d\omega'}{2\pi} J_3(\omega') I_{dc}(\omega').
 \end{equation}
 If one knows the dc characteristics $I_{dc}(\omega')$, one could in principle make a deconvolution to obtain $J_3(\omega')$; since $\omega_J=0$ is fixed, one needs an additional variable. \\
The simplest choice for the dc current would be a symmetric resonant structure at an energy $\hbar\omega_{res}$ varied, for instance, through a gate voltage:
 \begin{equation}
 I_{dc}(\omega')=2\pi Q_0\left[\delta(\omega'-\omega_{res})-\delta(\omega'+\omega_{res})\right],
 \end{equation}
 where $Q_0$ is a constant. Then one could have a direct access to the third cumulant through the photo-drag current in Eq.(\ref{photo_J3}): $$ 2iJ_3(\omega'=\omega_{res})=\frac{I_{f_{ac}}^{(0)}(\omega_J=0)}{Q_0}.$$
We will discuss in separate papers the possibility of this resonant behavior and the extension of such a method to cumulants of a phase operator $\hat{\varphi}_{ac}(t)$. \citep{ines_eugene_detection}
\label{sec_profiles}
\section{Application to a power law dc characteristics}
 \label{sec_power_law}
We will, through this section, explicit the formal results obtained previously for a gaussian and a lorentzian pulse in the case of a power law behavior of the dc current $I_{dc}(\omega_J)$, with an exponent $\alpha$. Assuming it's odd, we write it at positive dc voltages:
 \begin{equation}\label{Total_I_dc_TLL}
I_{dc}(\omega_J)=\left(\frac{\omega_J}{\omega_{c}}\right)^{\alpha}I_{\infty}^{(\alpha)}.
\end{equation}
 Here $\omega_c$ is a frequency cutoff and $I_{\infty}^{(\alpha)}$ is the value of the current at this cutoff, which depends on the exponent $\alpha$ and is of second order with respect to the operator $\hat{A}$ (see Eq.(\ref{Hamiltonian})). Then the differential conductance :
\begin{equation}\label{G_dc_TLL}
G_{dc}(\omega_J)=\frac{\alpha q}{\hbar \omega_{c}}\left(\frac{\omega_J}{\omega_{c}}\right)^{\alpha-1}I_{\infty}^ {(\alpha)},
\end{equation}
  has the same sign as $\alpha$. This is relevant to the impurity problem in a TLL with interaction parameter $K$. We focus on the weak-backscattering regime, for which $\alpha=2K-1$, so that $I_{dc}(\omega_J)$ in Eq.(\ref{G_dc_TLL}) corresponds to the backscattering current, reducing the perfect current $G_0V_{dc}$:
\begin{equation}\label{I_total_TLL}
I_{total}(\omega_J)=G_0V_{dc}-I_{dc}(\omega_J).
\end{equation}
We address the dual strong backscattering regime (or tunneling barrier) in appendix \ref{app_TLL}.  We also focus here on $K<1$, thus $-1<\alpha<1$. In a non-chiral wire with repulsive interactions, or equivalently, a coherent conductor connected to a resistance $R$, for which $1/K=1+R e^2h$,\citep{ines_saleur} one has $q=e$ and $G_0=e^2/h$. In the FQHE at a simple fractional filling factor (Laughlin series) $\nu=1/(2n+1)$ with integer $n$, one has $K=\nu<1/2$, thus $-1<\alpha<0$, $q=\nu e$, and $G_0=\nu e^2/h$.  
  
 One gets Eq.(\ref{Total_I_dc_TLL}) in the limit $k_B T\ll \hbar\omega_J$, but arbitrary temperatures could be considered as well (the theory holds even without an initial thermal distribution). In principle, periodic or non-periodic profiles of $f(t)$ could be implemented in Eq.(\ref{zero_frequency}) in order to obtain the current induced by $f(t)$. The arguments of $I_{dc}$ in the integral must be within the perturbative domain, which gives limitations on the profiles of $f(t)$ or energy scales one can use.\\
 As $\alpha<1$, requiring $|G_{dc}(\omega_J)|\ll G_0$ imposes a lower bound which defines the perturbative domain:
\begin{equation}\label{criteria}
\omega_J>\omega_B\simeq \omega_c |I_{\infty}^{(\alpha)}| ^{\frac{-2}{\alpha}}.
\end{equation}
 Thus the transfer rate function $p(\omega')$ in the integrand of Eq.(\ref{zero_frequency_ac}) must have its support at:
\begin{equation}\label{criteria_frequency}
|\omega'+\omega_J|>\omega_B. 
\end{equation} 
 Non-periodicity of $f(t)$ can make it easier to reach this criteria, as we illustrate it through the pulses considered in the previous section, the blorentzian (Eqs.(\ref{lorentzian},\ref{F_z_lorentzian})) or the gaussian one (Eq.(\ref{f_peak})). 
  \subsection{A gaussian pulse}
  We consider now $p(\omega)$ peaked around a frequency $\omega_p$, such as Eq.(\ref{f_peak}) with a small enough $\sigma$. One gets, from Eq.(\ref{zero-frequency-current_peak}):
  \begin{eqnarray}\label{current_peak}
I_f^{(0)}(\omega_J)&=&\left(\frac{\omega_J+\omega_p}{\omega_{c}}\right)^{\alpha}I_{\infty}^{(\alpha)}.
\end{eqnarray}
For an undetermined dc characteristics, we have already discussed the advantage of this relation to probe it in the out-equilibrium domain. Indeed, it is precisely in the latter, that for high voltages compared to temperature, that the power law in (\ref{Total_I_dc_TLL}) holds. One can choose vanishing dc voltage, to avoid heating for instance, but high $\omega_p\gg k_BT/\hbar$, so that the power law is probed through the photo-drag current $I_f^{(0)}(\omega_J=0)$ with respect to $\omega_p$.\\
A similar advantage holds if the dc voltage is too low to be inside the perturbative domain, defined by Eq.(\ref{criteria}), thus if $0<\omega_J<\omega_B$. One can rather increase $\omega_p$ to get $\omega_J+\omega_p>\omega_B$.   
\subsection{A Lorentzian pulse}
Consider now the lorentzian pulse with width $2\tau_1$ given by Eq.(\ref{lorentzian}). It induces the current in Eqs.(\ref{F_z_lorentzian},\ref{F_z_lorentzian_ac}), independently on the dc characteristics, replaced now by the power law in Eq.(\ref{Total_I_dc_TLL}). We subtract the dc current at the same $V_{dc}$, and rescale by its value at $1/2\tau_1$: 
\begin{eqnarray}\label{ratio}
\frac{I_f^{(0)}(\omega_J)-{I_{dc}(\omega_J)}}{I_{dc}\left(\frac{1}{2\tau_1}\right)}\!&=&\!2\alpha\tau_1\Omega_0 e^{2\tau_1\omega_J}\Gamma\left(\alpha,2\tau_1\omega_J\right).
\end{eqnarray} $\Gamma$ is the incomplete gamma function, and $\Omega_0=2\pi/T_0$, $T_0$ being the measurement time. This difference has the sign of $\alpha$, thus adding a pulse to $V_{dc}$ increases (resp. reduces) the current when $\alpha>0$ ($\alpha<0$). This can be explained by the fact that the dc current increases (resp. decreases) by increasing $V_{dc}$. As a consequence, in the TLL, the total current in Eq.(\ref{I_total_TLL}) is reduced (resp. increased) by the pulse when $K>1/2$ (resp. $K<1/2$). But now the dc average $I_{tot}(\omega_J)$ always increases with $\omega_J$, as $I_{dc}(\omega_J)$ is only a small correction to $G_0 V_{dc}$ and $G_0>0$. Thus the reduction of the total current due to the pulse for $K>1/2$ is rather counterintuitive. 

\begin{figure}[htb]\begin{center}
\includegraphics[width=6cm]{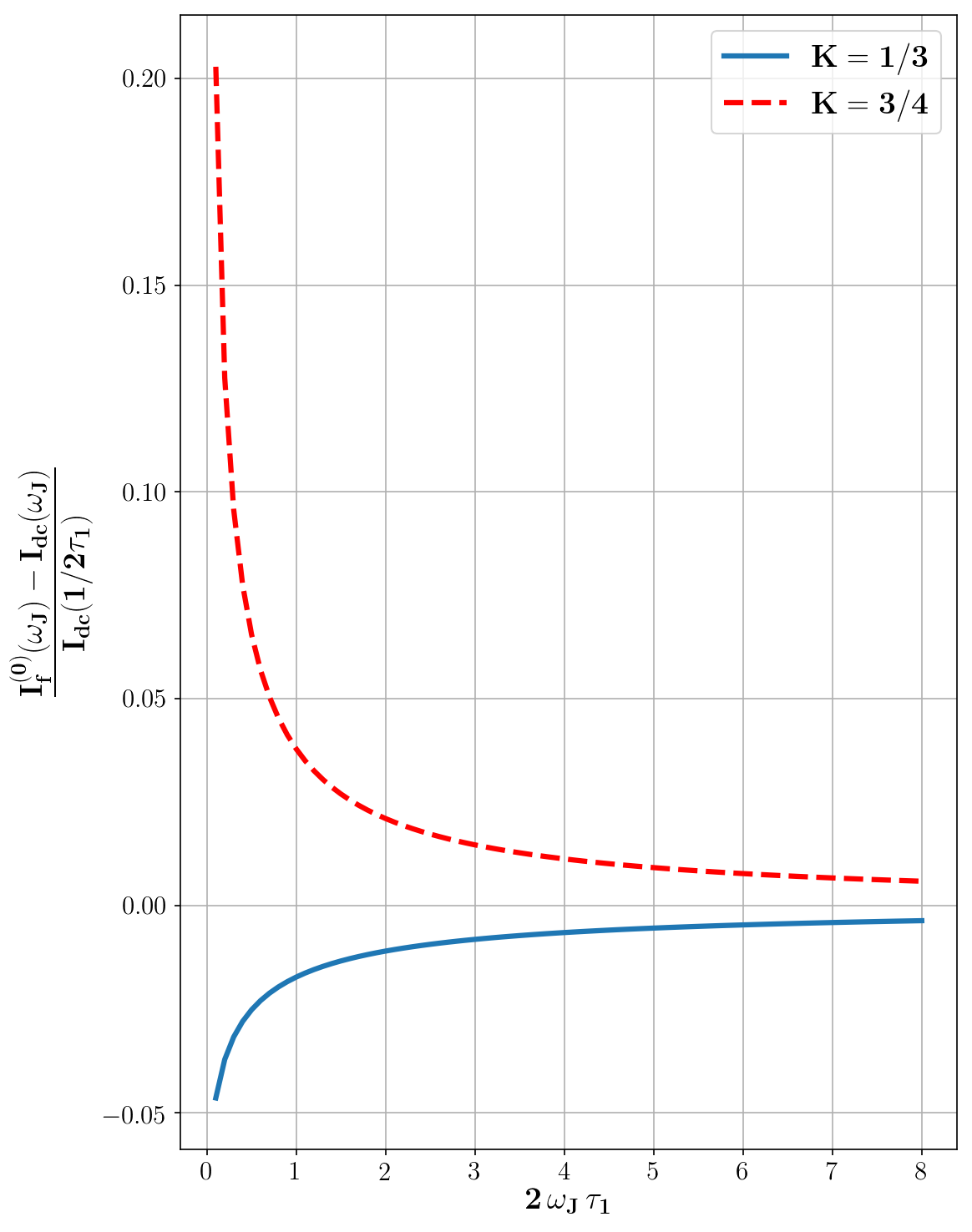}
\caption{\small  Weak impurity in a TLL with $K=1/3,3/4$, subject to a single lorentzian pulse with width $2\tau_1$ superimposed on a dc drive $\omega_J=qV_{dc}/\hbar$. The backscattering current is enhanced (resp. reduced) for $K>1/2$ (resp. $K<1/2$).}\label{fig_current_lorenztian_TLL_back}
\end{center}
\end{figure}

In Fig.\ref{fig_current_lorenztian_TLL_back}, we have plotted Eq.(\ref{ratio}) as a function of the dimensionless variable $2\tau_1\omega_J$, implicitly above $2\tau_1\omega_{B}$ (Eq.(\ref{criteria_frequency})). We choose a small width of the pulse compared to $T_0$, $2\tau_1\Omega_0=0.1$, and two values $K=1/3<1/2$ and $K=3/4>1/2$ for which the lorentzian pulse reduces (resp. increases) the backscattering current.
    In appendix \ref{app_TLL}, we show that the photo-conductance (see Eq.(\ref{photo_conductance})), gets reduced by the pulse for $0<\alpha<1$: $0\leq G_f(\omega_J)<G_{dc}(\omega_J)$, even though $I_{dc}(\omega_J)$ increases with $\omega_J$. The total photo-conductance in the TLL nonetheless increases. We also discuss the photo-drag current, which can probe the fractional charge and the power law behavior with respect to $\tau_1$. 
  \section{Determination of the fractional charge}
  \label{sec_charge}
   Here we discuss some methods of charge determination provided by the perturbative theory. For generality, we have introduced two renormalizing charges $q$ and $q'$ into Eq.(\ref{equivalence}) and Eq.(\ref{current}), which could be effective, determined for instance by $\mathcal{H}_0$. 
   
   On the one hand, in case $\hat{\rho}_0$ is a thermal density matrix, we have shown\citep{ines_cond_mat,ines_degiovanni_2016} that the shot-noise $S_{dc}(\omega_J)$ at high dc voltages $\hbar\omega_J\gg k_BT$ is universally poissonian, $S_{dc}(\omega_J)=q'I_{dc}(\omega_J)$, which provides $q'$.  \\
   On the other hand, only $q$ appears in $\mathcal{H}_{\hat{A}}(t)$ (unspecified, in Eq.(\ref{Hamiltonian})) as a parameter which links the Josephson-type frequency to the dc part of the voltage, $\omega_J=qV_{dc}/\hbar$. Then the robustness of this relation, through which $V_{dc}$ enters in Eq.(\ref{zero_frequency}), gives access to the charge $q$, independently on the underlying many-body eigenstates.   
   
   This is especially interesting for the FQHE with series of filling factors without a well-established Hamiltonian (in Eq.(\ref{Hamiltonian})), and with non-universal features such as non-abrupt edges, inhomogeneous filling factors or Coulomb interactions between edges. The fractional charge $q$ determining $\omega_J$ is then a parameter which depends on the dominant process at the quantum point contact, and the coupling to the TD voltage.
Compared to the poisson shot noise, which requires a large dc bias $qV_{dc}\gg k_B T$, thus can induce heating, our methods have the advantage to be based on the easier current measurement, and to be adapted to both weak $V_{dc}$ and $V_{ac}(t)$, which prevents heating.\citep{ines_eugene} The robustness of the fractional Josephson frequency $\omega_J=qV_{dc}/\hbar$ provides promising methods to determine the charge $q$, independently on the underlying Hamiltonian (obeying the minimal conditions).
For a periodic $f(t)$ at frequency $\Omega_0$, addressed in Ref.\citep{ines_eugene}, one method consists into exploiting the frequency locking, especially that zero-bias anomaly yields singularities in the differential conductance $G_{dc}(\omega_J+n\Omega_0)$ (indeed in its derivative). For non-periodic $f(t)$ addressed here, we have clarified and extended the lateral side-band transmission picture to many-body eigenstates exchanging continuous energy $\hbar\omega'$, so that $V_{dc}$ intervenes through the argument $G_{dc}(\omega_J+\omega')$. Thus robustness of the relation $\omega_J=qV_{dc}/\hbar$ still holds, and one can extend our previous methods\citep{ines_eugene} to a non-periodic $f(t)$.
 
Nonetheless, the present work provides two additional advantages.
On the one hand, we were not aware, in Refs.\citep{ines_eugene,ines_cond_mat}, that the relations we have obtained respectively for the average current and noise were indeed independent on the initial diagonal elements of $\hat{\rho}_0$. This offers an additional advantage compared to the poisson shot-noise, rather based on a thermal density matrix, as edges can lack thermalization between them or with the contacts. \\
   On the other hand, non-periodicity of $p(\omega)$ facilitates charge determination, as we illustrate it now through the simple peaked pulse at $\omega_p$, such as the gaussian one in Eq.(\ref{f_peak}) for small $\sigma$. This leads to Eq.(\ref{zero-frequency-current_peak}), a relation which goes well beyond the TLL model, and should apply to the FQHE at arbitrary series of filling factors, with possible Coulomb interactions between edges and different initial temperatures. We see that the r.h.s. of Eq.(\ref{zero-frequency-current_peak}) singles out a unique combination of the fractional $\omega_J$ with $\omega_p$. One can then proceed through two methods, depending on the freedom to vary respectively $\omega_p$ or $V_{dc}$. By plotting both members of Eq.(\ref{zero-frequency-current_peak}) as functions of $\omega_p$ or $V_{dc}$ respectively, the charge $q$ is guessed as a scaling parameter which ensures the equality.
   The first method, fixing $V_{dc}$ and varying $\omega_p$, is even easier when one lets $V_{dc}=0$, as Eq.(\ref{zero-frequency-current_peak_zero}) yields the photo-drag current $I_{f}^{(0)}(V_{dc}=0)\!\! =\!\! I_{dc}(V_{dc}=\hbar\omega_p/q)$. Thus $q$ scales $\omega_p$ in $ I_{dc}$ to get the observed dependence of $I_{f}^{(0)}(V_{dc}=0)$ on $\omega_p$. 
   The second method, fixing $\omega_p$ and varying $V_{dc}$, seeks to get the good scaling charge which enters through $\omega_J=qV_{dc}/\hbar$. \\
 One gains another advantage if the perturbative theory is defined by a lower bound, similar to Eq.(\ref{criteria}) obtained at simple filling factors. One can choose small $V_{dc}$, outside the perturbative domain, provided the frequency $\omega_p$ is high enough so that $\omega_p+\omega_J$ obeys Eq.(\ref{criteria}).
 \section{Conditions for the theory}
\label{sec_conditions}
Starting from the Hamiltonian (\ref{Hamiltonian}), the theory requires merely two conditions, in addition to Eq.(\ref{condition_matrix}). \\
First, we need, as it must, a weak operator $\hat{A}$, with respect to which second-order perturbation is valid and gives a finite result. \\
Secondly, we require the following cancellation: 
\begin{eqnarray}
\langle \hat{A}_0(t)\hat{A}_0(0) \rangle\!\!  &=&\!\!  0,\label{condition} 
\end{eqnarray}  
where the subscript $0$ refers to the interaction representation with respect to the Hamiltonian $\mathcal{H}_0$, which, for a given operator $\mathcal{B}$, is given by:
\begin{equation} 
	\mathcal{B}_0(t)\!\!  =\!\! 
	e^{\mi\mathcal{H}_0 t} \mathcal{B}\,e^{-\mi\mathcal{H}_0 t},
	\label{eq:heisenberg} 
\end{equation} 
Indeed, one should cancel the first order term in $C_f$, thus:
\begin{equation} \langle \hat{A}_0(t)\rangle\!\! =0.\label{condition_1}
\end{equation} But even not, the theory is still valid for the second-order term of $C_f$.\\
We can now formulate a sufficient condition to get Eqs.(\ref{condition},\ref{condition_1}), which makes it easier to distinguish a family of models described by the theory. For that, it is convenient to assume that a vertex operator enters into $\hat{A}$. Allowing for non-local processes, we use a multidimensional vector ${\bf x}$ to denote simultaneously various discrete entities and continuous spatial vectors within a region $\mathcal{C}$, and a phase field $\hat{\varphi}({\bf x})$:
 \begin{equation}\label{A(x)_0}
\hat{A}= \int_{\mathcal{C}} \! d{\bf{x}} \;e^{-i\hat{\varphi}({\bf x})}\;\hat{A}'(\mathbf{x}),
\end{equation}
Then $\varphi(t)$ (see Eq.(\ref{f_phase})) could correspond, fully or partly, to a position-independent average of a TD phase field (a kind of adiabatic approximation), with a fluctuating part given
by $\hat{\varphi}({\bf x})$, whose TD is due only to the Hamiltonian dynamics (such as in Eq.(\ref{eq:heisenberg})).\\
In case $\hat{\varphi}(\mathbf{x})$ and $\hat{A}'(\mathbf{y})$ commute, the sufficient condition ensuring Eq.(\ref{condition}) reads:
 \begin{equation} \label{invariance}
 \mathcal{S}_0 (\hat{{\varphi}}+c)\! =\!\mathcal{S}_0 (\hat{\varphi})
\end{equation}
for any real $c$, where $\mathcal{S}_0$ is the action associated with $\mathcal{H}_0$, and $\hat{\varphi}$ refers rather to its associated function of time, which is a global $U(1)$ gauge symmetry with respect to $\hat{\varphi}$. For instance, it holds whenever  $\mathcal{S}_0$ depends on ${\varphi}$ only through its time or space derivatives. As explained in appendix \ref{app_condition}, such an hypothesis is stronger than condition in Eq.(\ref{condition}). For many independent phase operators (see for instance Eqs.(\ref{A(x)_many_phases},\ref{Ai_vertex})), associated for instance with many elements of a quantum circuit or quantum channels, gauge invariance with respect to one among them is sufficient. \\We now notice two interesting consequences of Eq.(\ref{invariance}).\\
First, if the TD forces can be implemented through translating $\hat{\varphi}\rightarrow \hat{\varphi}+\varphi(t)$, the invariance of $\mathcal{S}_0$ in Eq.(\ref{invariance}) guarantees systematically that $\mathcal{H}_0$ does not acquire any time dependence, thus only the perturbing Hamiltonian $\mathcal{H}_{\hat{A}}(t)$ in Eq.(\ref{Hamiltonian}) depends on time.

Second, we can show, using Eq.(\ref{invariance}), that the dc average vanishes at $\omega_J=0$ to all orders with respect to $\hat{A}$, i.e. equation (\ref{vanishing}) becomes non-perturbative.  
 
\section{Unified examples}
\label{sec_examples}
 Here, we choose to keep a certain degree of generality and synthesis by writing a form for $\mathcal{H}(t)$ which is common to many physical problems with $N$ "entities" labelled by $i=1...N$. Referring to elements of a quantum circuit, electrodes, channels or edge modes in the integer, fractional or spin Hall effect, they are described by commuting Hamiltonians $\mathcal{H}_i$. We might add coupling terms $\mathcal{H}_{i,j}$, for instance Coulomb interactions between electrodes in a junction or between edge modes in the Hall regime, so that the Hamiltonian $\mathcal{H}_0$ in Eq.(\ref{Hamiltonian}) reads:
\begin{equation}\label{split_H0}
\mathcal{H}_0=\sum_{i=0}^N\mathcal{H}_i+\sum_{i\neq j}\mathcal{H}_{i,j}.
\end{equation} 
Let's provide now an example of a diagonal initial density matrix obeying Eq.(\ref{condition_matrix}). We can describe the case when the entities have different initial temperatures, with inverse values $\beta_1,... \beta_N$. But we need to assume $[\mathcal{H}_{k},\mathcal{H}_{i,j}]=0$ for all indices $i,j,k$; this holds in particular at $\mathcal{H}_{i,j}=0$, when mutual couplings are ensured only by $\hat{A}$. Then the initial global density matrix:
\begin{equation}\label{factorize}
\hat{\rho}_0=\frac{1}{\mathcal{Z}}\prod_{i}e^{-\beta_i \mathcal{H}_i},
\end{equation}
where $\mathcal{Z}=Tr \left[\prod_{i}e^{-\beta_i \mathcal{H}_i}\right]$, verifies the commutation relation in Eq.(\ref{condition_matrix}). 

Now, independently on the choice of $\hat{\rho}_0$, we choose to factorize $\hat{A}$ into mutually commuting operators $\hat{A}_i$ associated with the $N$ entities, dropping any space dependence in Eq.(\ref{A(x)_0}) for simplicity:
\begin{equation}
\label{A(x)_many_phases}
\hat{A}=\Gamma\prod_{i=1}^N \hat{A}_i^{\epsilon_i}.
\end{equation}
Here  $\Gamma$ is a complex amplitude, and $\epsilon_i=-1,1$ for $ \hat{A}_i$, $\hat{A}_i^{\dagger}$ respectively, while $\epsilon_i=0$ in absence of $\hat{A}_i$. 
Then ${\hat{A}}$ induces couplings between the $N$ entities of the system, in addition to those through $\mathcal{H}_{ij}$, with a different nature. \\
In many situations, $\hat{A}_i$ corresponds to a vertex operator expressed in terms of a phase operator $\hat{\varphi}_i$ and a real number $\alpha_i$:
 \begin{equation}\label{Ai_vertex}
 \hat{A}_i= e^{i\alpha_i\hat{\varphi}_i}.
 \end{equation}
Then $\hat{A}$ is a vertex associated with the total phase operator:
 \begin{eqnarray}
 \hat{\varphi}&=& \sum_{i=1}^N \epsilon_i{\alpha}_i{\hat{\varphi}}_i\label{total_phi}
\end{eqnarray}
Let's also introduce $\hat{\pi}_i$ (not necessarily a charge operator), the momentum conjugate to $\hat{\varphi}_i$: $[\hat{\pi}_i,\hat{\varphi}_i]=i$, so that $\hat{A}_i^{\epsilon_i}$ in Eq.(\ref{A(x)_many_phases}) translates $\hat{\pi}_i$ by $\epsilon_i \alpha_i$. Each Hamiltonian term $\mathcal{H}_i$ is a functional of $\hat{\varphi}_i$ and $\hat{\pi}_i$, as well on other operators. \\The invariance condition in Eq.(\ref{invariance}) has to be ensured by either $\hat\varphi$ in Eq.(\ref{total_phi}) or by a unique phase operator we choose, by permuting the indices, to be $\hat{\varphi}_1$.\citep{note_cond}

A particular case where the invariance becomes trivial is for $\mathcal{S}_0$ independent on $\hat{\varphi}_{1}$. Then the dynamics of $\hat{\pi}_{1} (t)$ is due only to $\mathcal{H}_{\hat{A}}(t)$ in Eq.(\ref{Hamiltonian}), and one can identify Eq.(\ref{eq:current}) with:
\begin{equation}
\hat{C}(t)=\frac{\epsilon_{1}}{\alpha_1}\partial_t \hat{\pi}_{1}(t).
\end{equation} 
In general, $\hat{\pi}_i$ or $\hat{\varphi}_i$ can be associated to operators of charge, spin current or voltages, for which the average of $\hat{C}(t)$ might as well provide a correction to their average.\\
We now specify to more typical systems for which $\mathcal{H}_0$ in Eq.(\ref{split_H0}) is relevant, with $\hat{A}$ given by Eq.(\ref{A(x)_0}) or Eq.(\ref{A(x)_many_phases}). 
\subsubsection{Normal, Josephson or Hall junctions}
A tunnel junction, coupling normal or superconducting conductors or edge states in the Quantum Hall regime, is a corner stone of transport studies, detection setups or density of states measurements, to mention only few examples. Even though they have facilitated explicit solutions for current average, previous studies of photo-assisted tunneling have often adopted the following restrictions: \begin{itemize}
 \item $\mathcal{H}_0$ (respectively $\hat{A}$) in Eq.(\ref{Hamiltonian}) is specified, split (respectively factorized) into operators associated with two independent electrodes and single particle states.\citep{tien_gordon,tucker_rev} 
 \item For superconducting electrodes, only quasiparticle current, with $q=e$ is considered, and not Josephson current. 
 \item A cosine ac voltage is typically considered, with a constant tunneling amplitude ($|{f}(t)|=1$).  
 \item The initial density matrix $\hat{\rho}_0$ is thermal.
 \item The inversion symmetry is adopted :
\begin{equation}\label{oddness}
I_{dc}(\omega_J)=-I_{dc}(-\omega_J).
\end{equation} 
\end{itemize}
Within our present theory, the perturbative relation in Eq.(\ref{zero_frequency}) is derived without any of those restrictions. We have seen that once specialized to a current operator, a cosine voltage, a constant modulus $|f(t)|=1$ and $q=e$, it reduces to Eq.(\ref{zero-frequency-periodic_charge}) with $p_l$ given by Eq.(\ref{bessel}), extending largely the validity of the Tien-Gordon theory. \citep{tien_gordon,tucker_rev}
Furthermore, our theory \citep{ines_eugene,ines_cond_mat} unifies the latter with two families of works devoted explicitly to barriers in the TLL, without any established connexion so far, and whose results are special cases of Eq.(\ref{zero-frequency-periodic_charge}): - a cosine ac voltage,\citep{wen_photo,sassetti_99_photo,photo_lin_fisher,photo_gogolin_TLL_PRL_93,photo_cheng_TLL_leads_06,photo_cheng_11,photo_crepieux,photo_TLL_ring_perfetto_13}  -
cosine modulated barriers.\citep{LL_pumping,photo_gefen_LL,enhanced_current_TLL_two_morais_PRB_06,enhanced_current_spin_hall_dolcini_2012,enhanced_current_TLL_one_imp_Zhou_PRB_06,enhanced_current_TLL_extended_morais_PRB_07,enhanced_current_komnik_TLL_one_radiative_TD_PRB_07,photo_spin_current_feldman_PRB_07,salvay} \\   In order to illustrate some of the additional features the theory could treat, we let  $N=2$ in Eq.(\ref{split_H0}) and assign $\mathcal{H}_1$ and $\mathcal{H}_2$ to two electrodes or sets of edge states in the integer, spin or fractional quantum Hall regime. Those don't reduce to kinetic terms, and can encode similar or different processes, such as Coulomb interactions or superconducting correlations. As we don't assume any initial thermalization, we could adopt the form in Eq.(\ref{factorize}) with different initial temperatures, which offers, to our knowledge, novel situations even within an independent electron framework. Even more, mutual finite-range and inhomogeneous Coulomb interactions are encoded within $\mathcal{H}_{12}$, which is quite realistic for Hall edges states at the quantum point contact (see Fig.\ref{setup_hall}). In addition, for filling factors of the FQHE which are not in the Laughlin series, microscopic descriptions are often subject to debates, so that keeping those Hamiltonian terms undetermined is extremely useful. \\
By allowing for a simultaneous and non-periodic TD of both $|f(t)|$ and $V_{ac}(t)$, one could, for instance, choose three different periods for $|f(t)|$ and $V_1(t),V_2(t)$ which could determine $V_{ac}(t)=V_1(t)-V_2(t)$. A time delay between $V_1(t)$ and $V_2(t)$ would allow to address Hang Ou Mandel (HOM) interferometry, whose results for noise are contained formally within our relations.\citep{ines_cond_mat}

 The operator $\hat{A}$ could couple many-body correlated states, or correspond to spatially extended tunneling, such as in Eq.(\ref{A(x)_0}), provided a unique effective TD holds through $f(t)$. 
We consider only local processes, adopting Eqs.(\ref{Ai_vertex},\ref{A(x)_many_phases}), with commuting $\hat{\varphi}_1,\hat{\varphi}_2$ associated with both electrodes or sets of edge states. Choosing $\alpha_{1}=-\alpha_2=1$, the Hamiltonian in Eq.(\ref{Hamiltonian}) reads:
   \begin{equation}\label{H_Hall}
   \mathcal{H}(t)=\mathcal{H}_1+\mathcal{H}_2+\mathcal{H}_{12}+f(t)\Gamma e^{i(\hat{\varphi}_1-\hat{\varphi}_2)}+h.c.
   \end{equation} 
 Now Eq.(\ref{total_phi}) reduces to:
 \begin{equation}
 \label{phi}
 \hat{\varphi}=\hat{\varphi}_1-\hat{\varphi}_2.
 \end{equation}
 Then the condition in Eq.(\ref{invariance}) has to be required  for only one phase operator among $\hat{\varphi}_{1}$, $\hat{\varphi}_{2}$ or $\hat{\varphi}$.
 
 Such phase operators and gauge invariance arise naturally within bosonisation, as $\mathcal{S}_0$ is a functional of time or space derivatives of those fields. $\hat{A}$ could refer to either strong or weak backscattering regimes, with a spatial extension in Eq.(\ref{A(x)_0}), or a superposition of various tunneling processes between many couples of edge states, with either a common charge and TD function $f(t)$ or with a unique dominant process. 
For helical edge states in quantum spin Hall insulators, $\hat{A}$ could describe extended umklapp processes, or other backscattering processes due to Coulomb interactions, and $\hat{C}(t)$ in Eq.(\ref{eq:current}) could be a spin current operator. 

The Hamiltonian in Eq.(\ref{H_Hall}) is relevant to a JJ, for which one has $q=2e$. Then $\hat{\varphi}$ in Eq.(\ref{phi}) corresponds to the Josephson phase, $\Gamma=E_J$ to a weak Josephson energy, and $\hat{C}(t)$ to the Josephson current operator. For the generic JJ, one has simply $\mathcal{H}_1+\mathcal{H}_2=E_c \hat{Q}^2/2$, where $E_c$ the charging energy, thus the gauge invariance in Eq.(\ref{invariance}) is trivial; but more complicated forms could be adopted. $E_J$ has to be small compared to $E_c$, or to an energy scale defined by the perturbative domain, which can be enlarged by coupling the JJ to an electromagnetic environment. 
\subsubsection{Strongly correlated tunnel or Josephson junctions in a quantum circuit}
A strongly correlated tunnel or Josephson junction such that as those addressed above can, in addition, be strongly coupled to an electromagnetic environment described by a third Hamiltonian term $\mathcal{H}_3$, letting $N=3$ in Eq.(\ref{split_H0}). This forms a strongly correlated quantum circuit, with simultaneous exchange of photons associated with the electromagnetic environment, the classical radiations and Coulomb interactions. The junction and the environment can have non-thermal initial distributions, or two different temperatures, as in Eq.(\ref{factorize}).\\
If the coupling terms between the two electrodes and the environment, $\mathcal{H}_{13} $ and $\mathcal{H}_{23}$, are linear, they can be absorbed through a gauge transformation, adding a phase field $\hat{\varphi}_3$ in $\hat{A}$, Eqs.(\ref{A(x)_many_phases},\ref{Ai_vertex}). In the non-local form, Eq.(\ref{A(x)_0}), $\hat{\varphi}(\bf{x})$ can be associated with the environmental degrees of freedom, and $\hat{A}'(\bf{x})$ with the electronic degrees of the freedom, but we focus again on local processes only. We denote by $\mathcal{H}_{el}=\mathcal{H}_1+\mathcal{H}_2+\mathcal{H}_{12}$ and $\hat{\varphi}_{el}=\hat{\varphi}_1-\hat{\varphi}_2$ (see Eq.(\ref{phi})), thus the total Hamiltonian in Eq.(\ref{Hamiltonian}) reads:
   \begin{equation}\label{H_Hall_3}
    \mathcal{H}(t)=\mathcal{H}_{el}+\mathcal{H}_{3}+f(t)\Gamma e^{i(\hat{\varphi}_{el}-\hat{\varphi}_3)}+h.c.
   \end{equation} 
It's now sufficient to ensure Eq.(\ref{invariance}) for either $\mathcal{H}_{el}$ or $\mathcal{H}_{3}$ with respect to $\hat{\varphi}_{el}$ or $\varphi_3$ respectively. In  Eq.(\ref{zero_frequency}), $I_{dc}(\omega)$ incorporates the effect of the environment. It is not odd for a non-gaussian environment, generating a photo-ratchet effect (because $\mathcal{H}_{3}(\hat{\varphi}_3)\neq \mathcal{H}_{3}(-\hat{\varphi}_3)$).\\
The Hamiltonian in Eq.(\ref{H_Hall_3}) applies as well to a JJ with a small $E_J$, coupled strongly to an electromagnetic environment, for which $\mathcal{H}_{el}=\mathcal{H}_{J}$ is the Hamiltonian of the isolated JJ. It describes also the opposite limit of high $E_J$, thus the dual phase-slip JJ, where current and voltage are permuted. $\hat{C}(t)=\hat{V}(t)$ corresponds now to a voltage operator, and an ac external periodic or non-periodic current is imposed through $W_{ac}(t)$. The relation in Eq.(\ref{zero_frequency}) links the average of $\hat{V}(t)$ for a TD current to its average for a constant current $I_0$, with $\omega_J=\pi I_0/e$. For a cosine current and a specific Hamiltonian of the phase-slip JJ, F. Hekking and collaborators\citep{photo_josephson_hekking} have computed explicitly the average of $\hat{V}(t)$, using Keldysh technique, then affirmed it to obey "the general theorem" in Refs.\citep{ines_cond_mat,ines_eugene} (i. e. Eq.(\ref{zero-frequency-periodic_charge}) with $p_l $ given by Bessel functions, Eq.(\ref{bessel})). 
 \section{Discussion}
Without recourse to Keldysh technique, a second-order S-matrix expansion allows us to use properties of the unperturbed Hamiltonian at equilibrium. This led us to extend the side-band transmission picture to a strongly correlated systems or quantum circuits, reflecting quantum superposition of many-body eigenstates of the unperturbed Hamiltonian. Such a collective quantum coherence was indeed underlying the plasmon scattering approach (within the inhomogeneous TLL). \citep{ines_schulz_group} It also emerges from the equivalence between the strong back-action of a dissipative environment and the microscopic impurity problem in the TLL.\citep{ines_saleur,nazarov_small_z,pierre_environment_11,ines_pierre,baranger_environment} We have established here more common features to the two problems of different nature; they hold through out-of-equilibrium relations, valid in a much larger context. The theory unifies quantum circuits and strongly correlated conductors either in 1-D, though special, with those in 2-D or 3-D. It allows potentially to understand the interplay between inelasticity, non-linearities and decoherence due, simultaneously, to Coulomb interactions, exchange of photons with radiations, and with an electromagnetic environment. \\The theory unifies also many works, based on different and explicit models, and restricted to periodic drives, such as the Tien-Gordon theory for photo-assisted tunneling,\citep{tien_gordon,tucker_rev,gabelli_08} impurities in the TLL,\citep{wen_photo,sassetti_99_photo,photo_lin_fisher,photo_cheng_TLL_leads_06,photo_cheng_11,salvay,photo_crepieux,photo_TLL_ring_perfetto_13,photo_spin_current_feldman_PRB_07} minimal excitations generated by lorentzian pulses.\citep{keeling_06_ivanov}  More recent works with two specific models of the phase slip JJ\citep{photo_josephson_hekking} and the TLL model (thus Laughlin series in the FQHE),\citep{martin_sassetti_prl_2017} are also special situations within our theory.\\

To our knowledge, inversion symmetry, thus an odd $I_{dc}(\omega_J)$, related often to particle-hole symmetry, is implicitly adopted in all those works. Then, for a cosine voltage, the photo-ratchet effect cannot be addressed,,\citep{tien_gordon,tucker_rev,gabelli_08} as $p_l$, the probability of exchanging $l$ photons, is symmetric: $p_l=p_{-l}$ (see Eqs.(\ref{p_periodic},\ref{zero-frequency-periodic_charge}). This symmetry is broken for the periodic lorentzian pulses, for which $p_lp_{-l}=0$. But L. Levitov {\it et al} do not consider an additional free dc drive $\omega_J$, and assume that the area of the pulse alone determines the current,\citep{keeling_06_ivanov} which holds only for a linear dc current, as shown in appendix \ref{app_linear}. The relation in Eq.(\ref{zero_frequency}) shows that $I_{f}^{(0)}(\omega_J)$ is rather determined in a non-trivial fashion by $f(t)$, even if we choose $|f(t)|=1$, and $V_{ac}(t)$ is formed by periodic or non-periodic series of lorentzian pulses. \\
We have shown here how an asymmetric $p(\omega)$ leads to a photo-ratchet effect, while breakdown of the inversion symmetry, thus asymmetric $G_{dc}(\omega_J)$, will be discussed in a separate paper. 

 Numerous other relations could indeed be derived for the noise and the generalized admittance of first\citep{ines_cond_mat,ines_eugene,ines_degiovanni_2016} or second order, and will be detailed in separate publications. Some have been tested experimentally within the Dynamical Coulomb Blockade for normal \citep{ines_portier_2015,hofheinz_photons_11,reulet} and JJs.\citep{hofheinz_photons_11,saliha} 
Within the FQHE with $\nu$ in the Jain series, subject a cosine voltage, the relation we have obtained for the photo-assisted noise\citep{ines_cond_mat} has ben exploited in Ref.[\citep{christian_photo_2018}] to determine $q$ through the  Josephson-type relation $\omega_J=qV_{dc}/\hbar$ (Fig.\ref{setup_hall}); we hope non-periodic profiles, more advantageous as illustrated here, could be implemented.

 Indeed, the fractional Josephson frequency has been already introduced through the photo-assisted current by X. G. Wen,\citep{wen_photo} but only for Laughlin series, thus $\nu=1/(2n+1)$ for which $q=\nu e$, and for edges described by the TLL model. Instead, we don't need here to know the underlying Hamiltonian for the edges, and all series could in principle be addressed. 
 
  Indeed, through the present study of $I_f^{(0)}(\omega_J)$, we have already the noise when the system is in its many-body ground state, and subject to one or series of periodic or non-periodic lorentzian pulses (for $|f(t)|=1$), as we have shown that: $S_f^{(0)}(\omega_J)=qI_f^{(0)}(\omega_J)$.\citep{ines_cond_mat} In particular, with a delay between two lorentzian series from the reservoirs, we have potentially the HOM noise, especially useful to explore the role of fractional statistics.

Compared to our work in Ref.\citep{ines_eugene}, we have shown that the initial density matrix has not to be thermal, but has only to commute with $\mathcal{H}_0$, Eq.(\ref{Hamiltonian}). We have given a sufficient condition in terms of a gauge invariance, and we have addressed the case of non-periodic or statistical mixture of coherent radiations, which led us to define properly the zero-frequency average of $\hat{C}(t)$ in view of possible singularities of $f(\omega)$.   
\section{Conclusion}
The paper gives an in-depth perturbative study\citep{ines_cond_mat} relevant to strongly correlated conductors or quantum circuits driven by coherent radiations, statistical mixture of radiations, or superposed ac voltages with different periods or delays superimposed on a free dc drive $\omega_J$. We define a weak operator which can refer to voltage, charge or spin current operators in Josephson, normal, magnetic junctions or edge states in the integer, fractional or spin Hall effects. Its out-of-equilibrium average is expressed as a continuous superposition of replicas of its average under a dc drive, sufficient to encode interaction effects. 
 This relation reflects quantum superposition of global many-body states exchanging photons at continuous frequencies $\omega$ with a transfer rate $p(\omega)$. It extends largely the lateral side-band transmission picture, usually restricted to a current operator, a cosine voltage and independent quasiparticles with charge $q=e$.\\ We have selected some applications for a charge current. First of all, the robustness of the frequency locking, where the dc voltage $V_{dc}$ intervenes only through the Josephson type frequency $\omega_J=qV_{dc}/\hbar$ with $q$ a charge parameter, offers various methods for determination of $q$ which are free from unknown parameters or microscopic descriptions. They are especially relevant to the FQHE, especially for filling factors $\nu$ beyond the simple fractions of the Laughlin series, $\nu\neq 1/(2n+1)$, still not well understood, and for non-universal features difficult to solve, such as mutual inhomogeneous interactions between the edges or their reconstruction. Our methods derived for periodic radiations\citep{ines_eugene} can be extended to non-periodic radiations or their statistical mixture, and to absence of initial thermalization. They are more convenient to use than poissonnian noise, as current is easier to measure than noise, and voltages can be low enough to avoid heating. Non-periodicity adds also advantages illustrated here.\citep{ines_cond_mat}  
 
 Secondly, the relations we have obtained allow one to infer the rectified current if both $p(\omega)$ and the dc current are known. That's the spirit of the application to the TLL, relevant to abrupt edge states in the FQHE at $\nu=1/(2n+1)$, one-dimensional interacting wire or a coherent conductor connected to a resistance.\citep{ines_saleur} A counterintuitive feature arises, questioning the terminology "photo-assisted": a lorentzian pulse superimposed on $V_{dc}$ reduces the current at the same $V_{dc}$, even when the dc current is an increasing function of $V_{dc}$ (thus for moderate repulsive interactions or resistance).

Thirdly, one can infer, inversely, the out-of-equilibrium dc current or the transfer rate $p(\omega)$ from the rectified current, or, operating at very low dc voltages to avoid heating, from the photo-drag current. The latter can be finite within our theory, contrary to previous works on photo-assisted current, as inversion symmetry and symmetry of $p(\omega)$ can be broken. This provides convenient spectroscopic methods of the out-of-equilibrium dc current on the one hand, by selecting $p(\omega)$ with its main weight in the out-of-equilibrium domain, and of the finite-frequency third cumulant of a statistical mixture of radiations (if weak) on the other hand.
 
More generally, the perturbative relations offer consistency tests of the underlying hypothesis of a given model, of numerical simulations or of experimental measurements. 
  They are especially relevant to two flourishing fields where interactions play a crucial role, and one seeks to generate and manipulate time-resolved excitations, or to explore individual or collective phenomena implying electrons and photons: - The electronic quantum optics, based mainly on the Integer Quantum Hall Effect with interacting edges,\citep{eugene_liste,plasmon_list,plasmon_mach_zehnder_kovrizhin_09,dubois_minimization_integer}  and on the FQHE, more difficult to manipulate -Quantum circuit electrodynamics, where a JJ or a quantum dot coupled to the modes of the electromagnetic environment simulate an atom-light interaction under control,\citep{circuit_electrodynamics_lukin_PRA_04,circuit_electrodynamics_kontos_PRL_11,circuit_electrodynamics_basset_PRB_13} and where our theory has proven to be relevant to squeezed states of photons.\citep{circuit_electrodynamics_mora_PRB_16,circuit_electrodynamics_mora_cea_PRB_17} 
It would be interesting to address more precisely conditions of relevance of the theory to quantum dots, or TD disorder and its interplay with correlations in cold atoms.\citep{photo_TD_disorder_giamarchi_PRB_06} 
\acknowledgments{The author dedicates this paper to an exceptional physicist and friend, Frank Hekking, who was so enthusiastic about this work, willing to read it few days before he left us. She is extremely grateful to B. Doucot for his suggestions and encouragements. She thanks D. C. Glattli, M. Kapfer, I. Taktak, P. Degiovanni, and R. Deblock for discussions and useful suggestions for the manuscript, and Q. Goutaland for technical help. She also thanks E. Sukhorukov and A. Borin for collaboration on the subject.}
\appendix
\section{A sufficient condition}
\label{app_condition}
Here we explain why the sufficient condition of gauge invariance with respect to one phase field, in Eq.(\ref{invariance}),  \begin{equation} \label{invariance_app}\mathcal{S}_0 (\hat{\varphi}+c)\!\!  =\!\! \mathcal{S}_0 (\hat{\varphi}),
\end{equation}
for any real $c$, leads to Eq.(\ref{condition}). $\mathcal{S}_0$ is the euclidian action associated with $\mathcal{H}_0$.
$\hat{\varphi}(\mathbf{x})$ appears in Eq.(\ref{A(x)_0}), and is assumed to commute with $\bar{\hat{A}}'(\mathbf{x})$. The phase argument $\hat{\varphi}$ of the action refers to the associated function of imaginary time and the multidimensional vector $\bf x$.
 Using the interaction representation in imaginary time with respect to $\mathcal{H}_0$ (see Eq.(\ref{eq:heisenberg}), we consider the m-vertex correlator:
\begin{equation}
X^{(m)}=\langle e^{-i a_1\hat{\varphi}_0(\bf{x_1},\tau_1)}.....e^{-i a_m\hat{\varphi}_0(\bf{x_m},\tau_m)}\rangle\!\! =\!\! 0,
\end{equation}
with $a_1,...,a_m$ real numbers. Writing its functional integral version, and translating the function $\hat{\varphi}_0((\bf{x},\tau)$ by an arbitrary constant $c$, we get $$X^{(m)} = e^{-\mi c \sum_{j=1}^m a_j} X^{(m)}.$$ Thus $X^{(m)}$ vanishes whenever $\sum_{j=1}^m a_j \neq 0$. Being valid for all integer numbers $m$ and many series of $a_i$, the condition (\ref{invariance_app}) is stronger than that in Eq.(\ref{condition}), to which it leads by letting $m=2$ and  $a_1=a_2=1$.  It leads as well to a first-order vanishing term, for $m=1$ and $a_1=1$ in Eq.(\ref{condition_1}).  \\ Equation (\ref{invariance_app}) is ensured when $\mathcal{S}_0$ depends only on gradients of first or higher orders of $\hat{\varphi}(\mathbf{x})$, as is the case in generic quantum circuits or bosonized models for which $\hat{\varphi}(\mathbf{x})$ is a bosonic field. \\
 \section{Formal derivation of the finite-frequency average}
\label{app_derivation}
Let's now give the main steps of derivation to relate formally the averages of the operator $\hat{C}(t)$ in Eqs.(\ref{Fdc},\ref{Fac_t}). We assume the energy scales used are within the perturbative domain, and conditions in Eq.(\ref{condition} and Eq.(\ref{condition_matrix}) hold. Then both averages, $C_{dc}(\omega_J)$ under a dc drive, in Eq.(\ref{Fdc}), and $C_{f}(\omega_J;t)$ under radiations, in Eq.(\ref{Fac_t}), can be expressed through a unique retarded correlation function:
\begin{eqnarray}\label{XR}
\hbar^2 X^R(t)\!\! &=&\!\! \theta(t)\left<\left[\hat{A}_0^{\dagger}(t),\hat{A}_0(0)\right]\right >,
\end{eqnarray}
where $\hat{A}_0(t)$ is given by Eq.(\ref{eq:heisenberg}) (see Eq.(\ref{Hamiltonian})). The correlators at different times depends only on their difference because $\mathcal{H}_0$ is independent on time, and commutes with $\hat{\rho}_0$ (Eq.(\ref{condition_matrix})).\\
Consider first the stationary regime, corresponding to $f(t)=1$. Then, $\langle \hat{C}_{\mathcal{H}}(t)\rangle= C_{dc}(\omega_J)$ in Eq.(\ref{Fdc}) is stationary (or at least its second-order term whenever Eq.(\ref{condition_1}) is not ensured), and is related to the Fourier transform of $X^R$:
\begin{equation}\label{I_DC}
C_{dc}(\omega_J)=2Re X^R(\omega_J).
\end{equation}
Let's now consider now the average under radiations, in Eq.(\ref{Fac_t}), where the subscript $f$ recalls its functional dependence on $f(t)$, while dependence on $\omega_J$ is more explicit. Its Fourier transform:
 \begin{equation}\label{Fac}{C}_{f}(\omega_J;\omega)=\int_{-\infty}^{+\infty} \!dt \;\;e^{\mi\omega t}
{C}_{f}(\omega_J;t),\end{equation} 
can be again expressed through $X^R$ in Eq.(\ref{XR}):
 \begin{eqnarray}\label{Iomega}
{C}_{f}(\omega_J;\omega)\! =\! \! \int \frac{d\omega'}{2\pi} {f}^*(\omega'-\omega/2)
f(\omega'+\omega/2)\nonumber\\\left[X^R(\omega_J+\omega'+\omega/2)+X^{{R^*}}(\omega_J+\omega'-\omega/2)\right].
\end{eqnarray}
We can deduce the zero-frequency limit, in Eq.(\ref{zero_frequency_w}), using Eq.~(\ref{I_DC}). For that, we need that $X^R(\omega)$ has no singularities at zero frequency, thus we require Eq.(\ref{supplementary_condition}). But we have kept $\omega$ on the r.h.s. of Eq.(\ref{zero_frequency_w}) in order to treat possible singularities of $f(\omega)$. Then ${C}_{f}(\omega_J;\omega\rightarrow 0)$ is expressed fully in terms of $f$ as well as $C_{dc}$. 

This statement can be extended to finite-frequencies. For that, we notice that one can, inversely, express $X^R$ in terms of $C_{dc}$, using Eq.~(\ref{I_DC}) and the Kramers-Kronig relation:
\begin{equation}\label{kramers}
2X^{R}(\omega)= C_{dc}(\omega)+i PP\int d\omega' \frac {C_{dc}(\omega')}{\omega'-\omega}.
\end{equation}
Upon substitution of Eq.(\ref{kramers}) into Eq.~(\ref{Iomega}), $C_{f}(\omega_J;\omega)$ becomes determined fully and universally by the function $C_{dc}(\omega)$. After some steps, we can cast it in the compact form in Eq.(\ref{Iomega_bis}), where  the Hamiltonian enters only through the out-of-equilibrium average $C_{dc}(\omega_J)$, obtained for $f(t)=1$.
We can check that in the stationary regime, one has $f(\omega)=2\pi\delta(\omega)$, so that, in view of Eq.(\ref{I_DC}), one has: ${C}_{f}(\omega_J;\omega)=2\pi\delta(\omega)C_{dc}(\omega_J)$. \\
 Using Eq.(\ref{zero_frequency}), and replacing $\hat{C}$ by a current for familiarity, one could express the differential photo-conductance $G_f(\omega_J)=dI_f^{(0)}(\omega_J)/dV_{dc}$ in terms of the differential dc conductance: $G_{dc}(\omega_J)={dI_{dc}(\omega_J)}/{dV_{dc}}$:
\begin{eqnarray}\label{photo_conductance}
G_f(\omega_J)&=&\left(|f_{dc}|^2+\frac{2}{T_0}{Re \left[f_{dc} f_{ac}^*(0)\right]}\right)G_{dc}(\omega_J)\nonumber\\&&+
\int_{-\infty}^{+\infty}\!\!\frac{d\omega'}{2\pi} p(\omega') G_{dc}(\omega_J+\omega').
\end{eqnarray}
   \section{A linear dc current}
 \label{app_linear}
We consider here that $\hat{C}(t)$ refers to a current, in Eq.(\ref{current}), whose dc average is linear: \begin{equation}\label{linear_F_dc}
 I_{dc}(\omega_{J})\simeq G_{dc}\;V_{dc},
 \end{equation}  
where $V_{dc}=\hbar\omega_{J}/q$ and $G_{dc}$ a linear dc conductance of second order with respect to $\hat{A}$, as is clear from the spectral decomposition in Eq.(\ref{spectral_F}). Linearity can still hold within strongly correlated systems, provided $\omega_{J}$ is within a certain (low-frequency) domain where $I_{dc}(\omega_{J})$ is analytic. For a thermal $\hat{\rho}_0$, this domain corresponds often to $\omega_{J}\ll k_B T/\hbar$, and $G_{dc}$ can then depend on temperature. \\
Now we use Eqs.(\ref{Iomega},\ref{I_DC}), replacing $C$ by $I$, and Eq.(\ref{linear_F_dc}). Using a general identity:
\begin{eqnarray}\nonumber
 \int\!\! \frac{d\omega'}{2\pi}{\omega'} {f}^*(\omega')f(\omega'+\omega) \!\!&=&\!\! \left[W \otimes \bar{|f^2|}-\frac{\omega}2\bar{|f^2|}\right]\!(\omega),
\end{eqnarray}
where $\otimes$ denotes the convolution product and $\bar{|{f}^2|}(\omega)$ the Fourier transform of $|{f}(t)|^2$, we obtain:
  \begin{equation}\label{average_linear}
 {I}_{f}(\omega_{J};\omega)=G_{dc}\bar{|{f}^2|}\otimes {V}(\omega),
 \end{equation}
 where $V(\omega)$ refers to the Fourier transform of the total voltage $V(t)$ in Eq.(\ref{V_dc_ac}).\\
Going back to time representation, the TD current average is a simple product:
 \begin{equation}\label{average_linear_time}
I_{f}(\omega_{J};t)=G_{dc} |{f}(t)|^2{V} (t).
 \end{equation}
 In particular, whenever $\phi_{ac}(t)$ (the phase of $f(t)$) is constant, and $V_{dc}=0$, thus $V(t)=0$, $I_{f}(\omega_{J}=0;t)=0$, even though the modulus of $f(t)$ varies in time. This feature is not generally valid for a non-linear $I_{dc}$. \\
 For a periodic $f(t)$, it is easy to convert the convolution in Eq.(\ref{average_linear}) into a discrete sum (see next appendix).
  For a non-periodic $f(t)$, one needs to take care of possible singularities in $\delta(\omega)$ when considering the zero-frequency limit. We decompose $f(t)$ in Eq.(\ref{dec_f}), $f(t)=f_{dc}+f_{ac}(t)$, $V(t)$ in Eq.(\ref{V_dc_ac}), and we adopt Eq.(\ref{g_meas}) where $T_0$ is the measurement time. Then the measured rectified current is given, in view of Eq.(\ref{average_linear_time}), by:
  \begin{eqnarray}\label{average_linear_meas}
 \frac{T_0}{G_{dc}}{I}_{f}^{(0)}(\omega_{J})&=&\left(|f_{dc}|^2T_0-2{Re(f_{dc} f_{ac}^*(\omega=0))}\right)V_{dc}\nonumber\\&&+\int_{\infty}^{\infty} dt |f_{ac}(t)|^2 {V}_{ac}(t),
 \end{eqnarray}
 Let us now specify further to $|{f}(t)|=1$, in which case equation (\ref{average_linear_time}) reduces to:
\begin{equation}\label{linear_I_e}
{I}_{f}(\omega_{J};\omega)=G_{dc}V(\omega),\end{equation}
thus is simply linear with respect to $V(\omega)$, as expected. Its dc measured component in Eq.(\ref{average_linear_meas}) reduces to:
\begin{equation}\label{linear_I_e_zero}
{I}_{f}^{(0)}(\omega_{J})=G_{dc}V^{(0)},
\end{equation}
 fully determined by $V^{(0)}$, the average of $V(t)$ over one period for a periodic $V(t)$, and $V^{(0)}=V_{dc}+V(\omega=0)/T_0$ (see Eq.(\ref{V0})) for non-periodic $V(t)$.
We see that for $|{f}(t)|=1$, the differentials of ${I}_{f}^{(0)}(\omega_{J})$ with respect to Fourier components $V_{ac}(\omega)$ vanish, but those are non-trivial when $I_{dc}(\omega_{J})$ is a non-linear function.\citep{ines_cond_mat,ines_eugene}
 \section{Periodic Driving}
\label{app_periodic}
Now we consider the case where $f(t)$ has a period $T_0=2\pi/\Omega_0$ (see also Ref.\citep{ines_eugene}). We keep the dc drive $\omega_{J}$ free, thus it has not to be commensurate with $\Omega_0$. The Fourier transform of $f(t)$ reads: \begin{equation}\label{periodic_e}
 f(\omega) = \sum_{l=-\infty}^{\infty} {f}_l \;\delta(\omega-l\Omega_0)
 \end{equation} 
We consider the average over one time period $T_0=2\pi/\Omega_0$ of $C_{f}(\omega_{\mathrm{dc}};t)$, in Eq.(\ref{Fac_t}), the analogous of Eq.(\ref{zero_frequency_ac}) for non-periodic TD:
\begin{equation}\label{zero-frequency-periodic}
C_{f}^{(0)}\;(\omega_{\mathrm{dc}})\!\! =\frac{1}{T_0}\int_0^{T_0}C_{f}(\omega_{\mathrm{dc}};t) \;dt = \!\! \sum_{l=-\infty}^{\infty}\! p_l \;C_{dc}(\omega_{\mathrm{dc}}+l\;\Omega_0)
\end{equation}
where $p_l=|{f}_l|^2$. 

At a finite frequency $\omega$, equation (\ref{Iomega_bis}) becomes:
\begin{eqnarray}\label{average_periodic}
C_{f}(\omega_{J};\omega)\!\!\!\!&=&\!\!\!\!\sum_{n=-\infty}^{\infty}   \delta(\omega-l\Omega_0)C_{f}^{(l)}(\omega_{J}),
\end{eqnarray} 
where $C_{f}^{(l)}(\omega_{J})$, not explicited here, is an integral implying $C_{dc}$ and the Fourier components ${f}_l$.\\
If now has also $\omega_{J}=l_{dc}\Omega_0$, then one can make the translation in the various sums:  $l\rightarrow l+l_{dc}$, which gives, for Eq.(\ref{zero-frequency-periodic}) for instance:
 \begin{equation}\label{zero-frequency-periodic_bis}
C_{f}^{(0)}(\omega_{\mathrm{dc}})\!\! =\!\! \sum_{l=-\infty}^{\infty}  p_{l-l_{dc}} C_{dc}(l\Omega_0).
\end{equation}
If we specify to a current operator, replacing $C$ by $I$ in Eq.(\ref{zero-frequency-periodic_bis}), we get also the transferred charge during one cycle:
\begin{eqnarray}\label{app_zero-frequency-periodic_charge}
Q(\omega_{\mathrm{dc}})\!\! &=&T_0I_{f}^{(0)}(\omega_{\mathrm{dc}})\nonumber\\
&&=T_0\!\! \sum_{l=-\infty}^{\infty}\! p_l \;I_{dc}(\omega_{\mathrm{dc}}+l\Omega_0).
\end{eqnarray}
For weak barriers with the same effective TD, this is nothing but the pumped charge. 
A similar relation would hold if a phase operator $\hat{\phi}$ determines $\hat{C}(t)=\partial_t\hat{\phi}(t)$, as is the case for a voltage operator in phase-slip JJs.\\

It is useful to write as well the photo-conductance in Eq.(\ref{photo_conductance}), where, instead of $\omega_{J}$, we use $V_{dc}$:
\begin{eqnarray}\label{photo_conductance_periodic}
G_f(V_{dc})\!\! &=&
\!\! \sum_{l=-\infty}^{\infty}\! p_l \;G_{dc}(V_{\mathrm{dc}}+l\hbar\Omega_0/q).
\end{eqnarray}
We stress that differentials with respect to $V_{ac}(\omega)$ are different.
 \subsubsection*{Periodic series of pulses: Josephson-type modulation}
It is worth recalling a very interesting application, performed without specifying the Hamiltonian, neither the diagonal initial density matrix$\hat{\rho}_0$. \citep{ines_eugene} We choose $|{f}(t)|=1$ and ${W}_{ac}(t)$ formed by a periodic series of pulses of area $\varphi$. For sharp pulses ${W}_{ac}(t)=\varphi\sum_l\delta(t-2\pi l/\Omega_0)$, we obtain: \citep{ines_eugene}
\begin{equation}\label{zn_pulses}
|{f}_{l}|^2=\left[\frac{\sin(\varphi/2)}{\pi l-\varphi/2}\right]^2.
\end{equation}
Without need to any superconducting type correlations, this leads, interestingly, to a Josephson type oscillating term in Eq.(\ref{zero-frequency-periodic}), $\sin^2(\varphi_0/2)$ :
\begin{equation}
C_{f}^{(0)}(\omega_{J})=\sin^2(\varphi/2)\sum_{l=-\infty}^{\infty}  \frac{4C_{dc}(\omega_{\mathrm{dc}}+l\Omega_0)}{(\varphi-2\pi l)^2}.
\end{equation}
One needs that the sum over $l$ on the r.h.s. doesn't have a peculiar behavior which would cancel the oscillations. For instance, if $C_{dc}$ has a power law behavior, such as is the case for the average current in a TLL model with parameter $K$, where $I_{dc}(\omega_{J})$ behaves as $\omega_{J}^{2K-1}$, the Josephson type oscillating function is preserved, and the series above converges.\\
 It would be interesting to address resonance conditions for either $\varphi$ or $\Omega_0$, for instance to determine $K$ or the fractional charge in the FQHE.
  \section{The impurity problem in a TLL-Application to a Lorentzian pulse}
  \label{app_TLL}
\begin{figure}[htb]\begin{center}
\includegraphics[width=6cm]{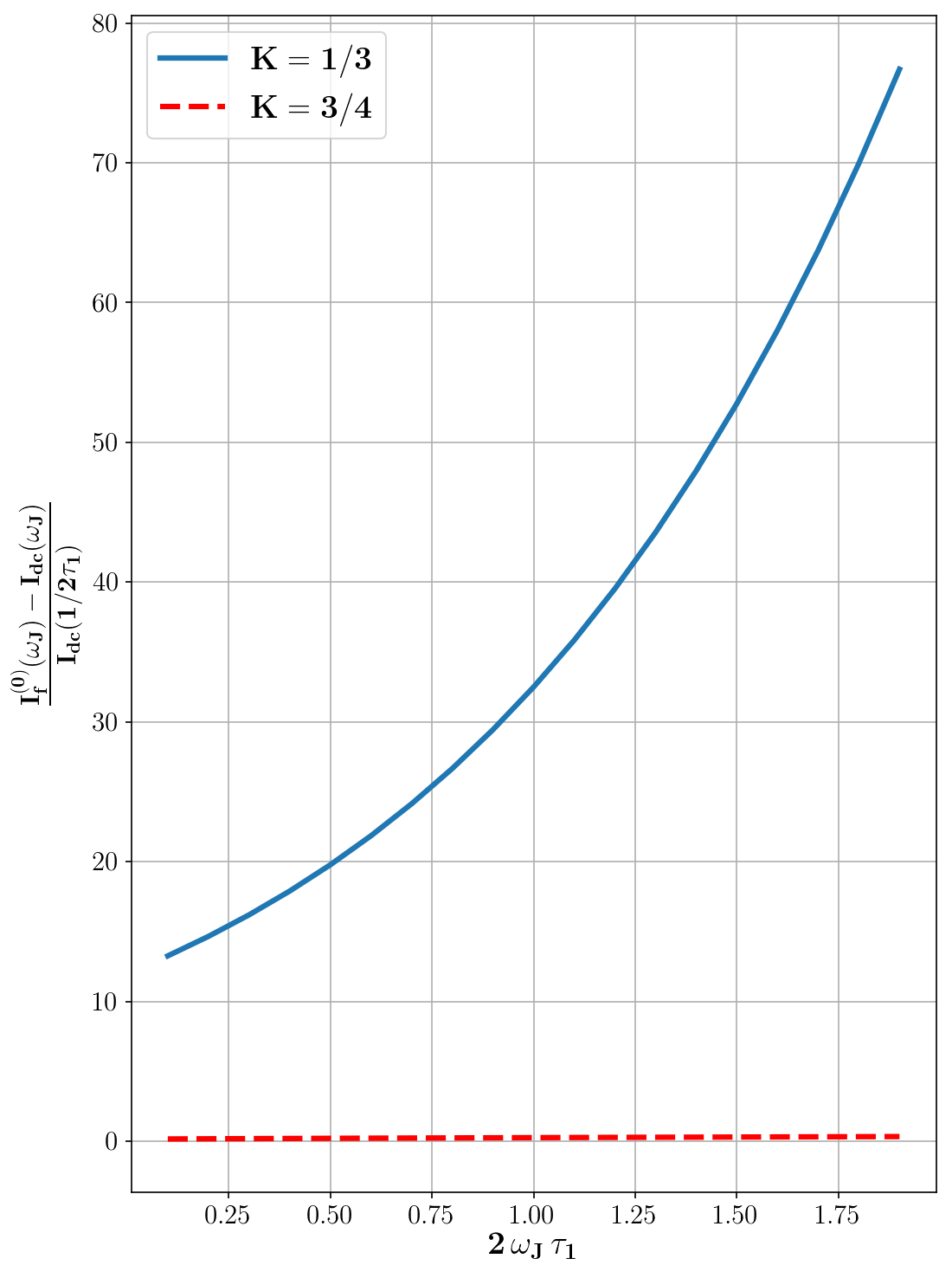}
\caption{\small The strong backscattering regime (or a tunneling barrier) in a TLL with $K=1/3,3/4$, subject to a lorentzian pulse with width $2\tau_1$ superimposed on a dc drive $\omega_J=qV_{dc}/\hbar$. The difference between the induced current $I_f(\omega_J)$ and the dc current $I_{dc}(\omega_J)$ at the same $\omega_J$, renormalized by $I_{dc}(\omega_J=1/2\tau_1)$,  is positive $K=1/3$, and almost vanishing for $K=3/4$.}\label{fig_current_lorenztian_TLL_tunn}
\end{center}
\end{figure}

 The TLL model, characterized by an interaction parameter $K$, turns out to be relevant to a large variety of systems, such as edge states in the FQHE or IQHE with interactions, spin Hall edge states of topological insulators, or quantum wires with reservoirs.  We have also shown its relevance to a coherent conductor connected to a resistive environment with resistance $R$ where $K=1/(1+r)$ and $r=e^2/h R$.\citep{ines_saleur} 
\\
For two edge states in the FQHE  with a constriction created by a gate, the TLL model is expected to be valid for an abrupt confinement of the edges and at simple filling factors $\nu=K=1/(2n+1)\leq 1/3$ with integer $n$. Then the power law in Eq.(\ref{Total_I_dc_TLL}) is obtained, provided tunneling between edges is local and mutual Coulomb interactions are neglected.\\ The typical Hamiltonian corresponds then, in Eq.(\ref{Hamiltonian}), to $\mathcal{H}_0$ given by the TLL Hamiltonian, or to Eq.(\ref{H_Hall}) where $\mathcal{H}_1$ and $\mathcal{H}_2$ are quadratic forms describing two chiral edge states. We don't explicit $\mathcal{H}_0$, which leads to the dc current already given by Eq.(\ref{Total_I_dc_TLL}). The operator $\hat{A}$ in Eq.(\ref{Hamiltonian}) describes one among two dual processes:
\begin{enumerate}
\item Weak-backscattering processes, for which :
\begin{equation}\label{alpha_weak}
 \alpha=2K-1.
 \end{equation} In this regime, $I_{dc}(\omega_J)$ in Eq.(\ref{Total_I_dc_TLL}) corresponds to the backscattering current, inducing a correction to the perfect linear current, (see Eq.(\ref{I_total_TLL})).
Whenever $K<1$, thus $\alpha-1<0$, the perturbative expression in Eq.(\ref{Total_I_dc_TLL}) is limited by Eq.(\ref{criteria}).\\ In the FQHE, one has tunneling of fractional charges between the upper and lower edges, $q=Ke$. Furthermore $K=\nu<1/2$, thus $\alpha<0$, so $I_{dc}(\omega_J)$ decreases with the dc voltage ($G_{dc}<0$, Eq.(\ref{G_dc_TLL})).
\item For the strong-backscattering regime, corresponding to $\omega_J<\omega_B$, one starts from weak tunneling of (in general) integer charges, $q=e$. The exponent in Eq.(\ref{Total_I_dc_TLL}) is now given by:
\begin{equation}\label{alpha_tunn}
 \alpha=2/K-1.
 \end{equation}  Then $I_{dc}(\omega_J)$ corresponds to a tunneling current, and perturbation is not restricted as above for $K<1$. For attractive interactions, thus $K>1$, one could have a criteria similar to Eq.(\ref{criteria}).
\end{enumerate}
We have applied the relation in Eq.(\ref{zero_frequency}) to the weak-backscattering regime, and to a lorentzian pulse, yielding the backscattering current in Fig.(\ref{fig_current_lorenztian_TLL_back}). In the dual regime, for which one has Eq.(\ref{alpha_tunn}), taking $K<1$ leads to $\alpha>0$, thus the pulse always increases, or "photo-assist" the current, as illustrated in Fig.(\ref{fig_current_lorenztian_TLL_tunn}).
Let's now express the photo-conductance in Eq.(\ref{photo_conductance}):
\begin{eqnarray}\label{conductance_lorentzian}
G_f(\omega_J)&=&G_{dc}(\omega_J)+2(\alpha-1)\tau_1\Omega_0\nonumber\\&&G_{dc}\!\!\left(\frac1{2\tau_1}\right)e^{2\tau_1\omega_J}\Gamma\left(\alpha-1,2\tau_1\omega_J\right),
\end{eqnarray}
where $G_{dc}(\omega_J)$ is given by Eq.(\ref{G_dc_TLL}). Since $G_f(\omega_J)$ and $G_{dc}(\omega_J)$ have both the sign of $\alpha$,\citep{note_sign} we get: \begin{equation}
|G_f(\omega_J)|<|G_{dc}(\omega_J)|.
\end{equation}
A counterintuitive feature arises when $0<\alpha<1$, as $I_{dc}(\omega_J)$ increases with $\omega_J$, while the pulse reduces the conductance: $0\leq G_f(\omega_J)<G_{dc}(\omega_J)$, which
questions the terminology of "photo-assisted" transport. But in the TLL, in the weak-backscattering regime, the pulse always increases the total conductance, in view of Eq.(\ref{I_total_TLL}), whenever $1>K>1/2$ or $K<1/2$.  \\

Let's now discuss the limit of a zero dc voltage, thus the photo-drag current. No caution is required for $\alpha-1>0$, but, for $\alpha-1<0$, $G_{dc}(\omega_J)$ diverges, 
 so that the limit $\omega_J=0$ cannot, strictly speaking, be undertaken into Eq.(\ref{conductance_lorentzian}), as one leaves the perturbative domain.\\
 Nevertheless, if one takes from the beginning $\omega_J=0$, so that the dc component of the voltage comes only from $V_{ac}(\omega=0)/T_0=V^{(0)}$ (see Eq.(\ref{V0})), the result for $G_f(\omega_J=0)$ is finite provided one chooses $1/2\tau_1$ within the perturbative domain, thus $2\tau_1\omega_B \gg 1$ (see Eq.(\ref{criteria})). 
Then, this amounts to ignore the terms depending on $\omega_J$ in Eq.(\ref{conductance_lorentzian}), which reduces to a power law with respect to $\tau_1$:
 \begin{equation}\label{G_0_TLL_Lor}
 G_f(\omega_J=0)=2\tau_1\Omega_0\Gamma(\alpha)G_{dc}(1/2\tau_1),
 \end{equation} 
 where $\Gamma(\alpha)$ is the Gamma function. 
In both regimes of weak or strong backscattering, this power law with respect to the width of the lorentzian pulse $2\tau_1$ can also provide a complementary test of such a behavior and method to measure $\alpha$, thus $K$ in the TLL. This is especially useful if the out-of-equilibrium regime is difficult to reach, as if high $V_{dc}$ causes heating.  

Similar lines as those for a peaked pulse in Eq.(\ref{f_peak}) could be followed as well for charge determination (see section \ref{sec_charge}), replacing $\omega_0$ by $1/2\tau_1$, and using the photo-drag current at $\omega_J=0$ in Eq.(\ref{F_z_lorentzian_current}). This is however less convenient due to integration performed on its r.h.s.. But in case of the TLL, thus for $\nu=1/(2n+1)$ in the FQHE, the expression in Eq.(\ref{F_z_lorentzian_current}), or its differential with respect to the dc drive, reduces to Eq.(\ref{G_0_TLL_Lor}). So one would need to vary the width of the pulse $2\tau_1$, and guess $q$ as a scaling factor between the frequency $1/2\tau_1$ and the natural argument of $G_{dc}$, the dc voltage.

\begin{thebibliography}{95}
\bibitem{note_revue}
The main powerful approaches thus adapted are the scattering approach of TD
  transport,\citep{buttiker_revue_time} Floquet theory \citep{photo_review} or
  the Tien-Gordon theory for photo-assisted tunneling,\citep{tien_gordon}
  extensively exploited for microwave astronomical detection or
  mixers.\citep{tucker_rev}.
\bibitem{note_dimension}
Notice that $C_{f}^{(0)}(\omega_J)$ has now the same dimension as the dc
  average $C_{dc}(\omega_J)$.
\bibitem{note_total_0}
Our discussion holds also if the regular dc part $W_{ac}(0)$ in Eq.(\ref{W0})
  vanishes, thus for a zero effective dc component, $W^{(0)}=0$.
\bibitem{note_cond}
One has to make sure that $\mathcal{H}_{1,i}$ does not induce a finite
  commutator $\left[\hat{\varphi}_1(t),\varphi_i(t')\right]$.
\bibitem{note_sign}
In particular, $G_f(\omega_J)<0$ for $\alpha<0$, which is again not so
  surprising as this is due to a negative dc conductance $G_{dc}(\omega_J)$.
  This remark was addressed in the case of a cosine weak barrier in the TLL
  with $K<1/2$, thus $\alpha=2K-1<0$, in the limit of vanishing dc
  voltage.\citep{photo_gefen_LL} Indeed, the relation we have obtained in
  Eq.(\ref{photo_conductance}) allows one to clarify it in a much larger
  context, as will be discussed in a separate paper. The issue we address here
  is different.
\end{thebibliography}

\end{document}